\documentclass[aip,jcp,reprint,twocolumn]{revtex4}
\usepackage{graphicx}
\usepackage{dcolumn}
\usepackage{bm}
\usepackage{graphicx}
\usepackage{dcolumn}
\usepackage{bm}
\usepackage{dcolumn}
\usepackage{amsmath}
\usepackage{graphicx}
\usepackage{dcolumn}
\usepackage{bm}
\usepackage{easyReview} 
\usepackage{multirow}
\usepackage{xcolor}
\usepackage[range-units=single,per-mode=reciprocal]{siunitx}

\newcommand{\ro}  { {\bf r}}
\usepackage[caption=false]{subfig}
\usepackage[normalem]{ulem}
\makeatletter
\def\squiggly{\bgroup \markoverwith{\textcolor{red}{\lower3.5\p@\hbox{\sixly \char58}}}\ULon}
\makeatother
\usepackage{graphicx}
\usepackage{dcolumn}
\usepackage{bm}
\usepackage{graphicx}
\usepackage{dcolumn}
\usepackage{bm}
\usepackage{dcolumn}
\usepackage{amsmath}
\usepackage{graphicx}
\usepackage{dcolumn}
\usepackage{bm}
\usepackage{easyReview} 
\usepackage{multirow}
\usepackage{xcolor}
\usepackage{easyReview}
\usepackage[range-units=single,per-mode=reciprocal]{siunitx}

\usepackage[caption=false]{subfig}
\usepackage[normalem]{ulem}

\makeatletter
\def\squiggly{\bgroup \markoverwith{\textcolor{red}{\lower3.5\p@\hbox{\sixly \char58}}}\ULon}
\usepackage{soul}
\makeatother
\begin{document}

	\title{Polymer Complexation: Partially Ionizable Asymmetric Polyelectrolytes }
	\author{Souradeep Ghosh$^{1,2}$}
	\author{Soumik Mitra$^{1}$}
	\author{Arindam Kundagrami$^{1,2}$}
	\altaffiliation{Corresponding author email: arindam@iiserkol.ac.in}
	\affiliation{$^{1}$Department of Physical Sciences, Indian Institute of Science Education and Research
		Kolkata, Mohanpur 741246, India}
		\affiliation{$^{2}$Centre for Advanced Functional Materials, Indian Institute of Science Education 
		and Research Kolkata, Mohanpur 741246, India}

	\begin{abstract}

Studies of the thermodynamics of complex coacervation of pairs of symmetric, strongly ionizable, oppositely charged polyelectrolyte chains are abundant. To generalize such understanding to asymmetric chain lengths and variable ionizability (chemical charge density), frequently observed in experiments, we present a theoretical framework to analyze the effective charge and size of the complex and the thermodynamics of complexation of two polyions as a function of such asymmetries. The free energy ensuing from the Edwards' Hamiltonian undergoes variational extremization, and explicitly accounts for the screened Coulomb and non-electrostatic interactions among monomers within individual polyions and between two polyions. Assuming maximal ion-pair formation of the complexed part, the system free energy comprising configurational entropy of the polyions and free-ion entropy of the small ions is minimized. The thermodynamic drive for complexation is found to increase with the ionizability of the symmetric polyions and to be maximum for symmetric chain lengths for equally ionizable polyions. The effective charge and size of the complex increase with asymmetry in charge density, where the size can be substantially larger than a collapsed globule found for symmetric chains. The regimes of enthalpy- and entropy-driven complexation are found, respectively, for low and high Coulomb strengths. The crossover strength is found to be strongly dependent on the dielectric environment and salt, but marginally dependent on the charge density, thus implying an entropy-driven process at moderate strengths. The key results match the trends in simulations and experiments, and are expected to provide insight for asymmetric complexation in real systems. 
 
	\end{abstract}

	\maketitle
	
	\section{Introduction}

The simple picture of complexation through ion pair formation between a pair of oppositely charged polyelectrolyte (PE) chains is now widely considered\cite{priftis2012,vitorazi2014,chang2017,adhikari2018,yethiraj2021,chen2022} as the primary step in the formation of phase separated complex coacervates. It is well established that though Coulomb attraction between the PEs prima facie seem to be the critical factor, it is actually the entropy of the released counterions that provides the major thermodynamic drive\cite{michaels1965,tianaka1980,dautzenberg2002,zhaoyang2006,gummel2007,larson2009,beltran2012,lemmers2012,tirrell2012,
gucht2012,semenov2012,perry2015,peng2015,dzubiella2016,salehi2016,fu2016,muthu2017,chang2017,radhakrishna2017,
lytle2017,schlenoff2017,meka2017,tirrell2018,adhikari2018,whitmer2018,whitmer2018macro,lytle2019,wang2019,
tirrell2020,tirrell2021,mitra2023} for the process. The Coulomb energy change of bound ion-pairs may even oppose complexation, and that too, counter intuitively, at higher electrostatic (Coulomb) strengths\cite{lundbck1996,matulis2000,bronich2001,zhaoyang2006,laugel2006,whitmer2018macro,
mitra2023}. Polyeletrolyte complexes (PEC) and coacervates are ubiquitous, and the thermodynamics of the formation of such structures is still a matter of intense discussion\cite{spruijt2013,perry2015,salehi2015,pablo2016,panyukov2018,zhang2018,wang2019,chen2022,lutkenhaus2015,fu2016,sing2017,
radhakrishna2017,michaels1965,tianaka1980,dautzenberg2002,zhaoyang2006,gummel2007,larson2009,beltran2012,lemmers2012,
tirrell2012,semenov2012,perry2015,peng2015,salehi2016,fu2016,muthu2017,lytle2017,schlenoff2017,meka2017,tirrell2018,
adhikari2018,whitmer2018,whitmer2018macro,wang2019,tirrell2020,tirrell2021,mitra2023,chen2022-PNAS}. 

There has been significant work in theory\cite{voorn1957,borue1988,borue1990,mahdi2000,joanny2001,delacruz2003,delacruz2004,kudlay2004,oskolkov2007,perry2015,
salehi2016,sing2017,lytle2017,adhikari2018,potemkin2017, panyukov2018,lytle2019,zhang2018,ylitalo2021,
wang-zhao2019,rumyantsev2018,rumyantsev2019,chen-yang2021,knoerdel2021,sayko2021,mitra2023,chen2022-PNAS}, simulation\cite{hayashi2002,winkler2002,hayashi2003,hayashi2004,zhang2005,zhaoyang2006,fredrickson2007,larson2009,semenov2012,
fredrickson2012,juarez2015,peng2015,lytle2016,dzubiella2016,fredrickson2017,chang2017,radhakrishna2017,whitmer2018,
whitmer2018macro,rumyantsev2019macro,shakya2020,neitzel2021,sayko2021,bobbili2022,chen2022,chen2022-PNAS} 
and experiments\cite{record1978,kabanov1985,dautzenberg2002,cousin2005,laugel2006,gummel2007,porcel2009,spruijt2010macro,chollakup2010,gucht2011,
tirrell2012,lemmers2012,gucht2012,tirrell2013,tirrell2014,vitorazi2014,perry2014,salehi2015,lutkenhaus2015,
kayitmazer2015,fu2016,meka2017,schlenoff2017,ali2018,dePablo2018,tirrell2018,ali2019,wang2019,huang2019,pde2020,
tirrell2020,meng2020,tirrell2020,tirrell2021,chen2021,neitzel2021,priftis2012,priftis2012-softmatter,subbotin2021,
friedowitz2021,ma2021,lalwani2021,digby2022} on solutions of PECs and phase coexistence of coacervates and supernatants therein. The entropy of the free counterions, however, manifests most significantly in the high dilution limits, but this regime offers exprerimental challenges\cite{pde2020}. Although several experiments\cite{michaels1965,tianaka1980,dautzenberg2002,gummel2007,beltran2012,lemmers2012,gucht2012,tirrell2012,tirrell2013,
tirrell2014,fu2016,meka2017,schlenoff2017,tirrell2018,wang2019,tirrell2020,tirrell2021} and simulations\cite{zhaoyang2006,larson2009,semenov2012,peng2015,dzubiella2016,chang2017,radhakrishna2017,whitmer2018,whitmer2018macro}, most of them recent, have stressed on the counterion entropy as the major thermodynamic drive for complexation, the drive is found to be enthalpic at low Coulomb strengths for implicit solvents\cite{zhaoyang2006,whitmer2018macro,mitra2023}. This interpretation is challenged once solvent degrees of freedom are considered in a polar solvent\cite{fu2016,schlenoff2017,wang2019,chen2022-PNAS}, for which the drive is found to be entropic even at low Coulomb strengths. Notably, this complex interplay of entropy and enthalpy in the presence of Coulomb screening is effective in the thermodynamics of complexation of charged biopolymers, such as protein-polyelectrolytes \cite{DeRouchey2005,cousin2005,gummel2007,daSilva2009,kayitmazer2013,yigit2015} and protein-protein\cite{hofmann2019,borgia2018} pairs, as well.
 
Starting with the seminal work of Voorn and Overbeek (VOT)\cite{voorn1957} that combined the mean-field Debye-H\"uckel\cite{mcquarrie2000} theory for small electrolytes along with the Flory-Huggins\cite{doi1996} mixing entropy, the PEC theories progressed through advanced treatments incorporating more accurate charge correlations like the random phase approximation(RPA)\cite{borue1988,borue1990,mahdi2000,joanny2001,delacruz2003,delacruz2004,kudlay2004,oskolkov2007,
potemkin2017,rumyantsev2018,chen-yang2021,sayko2021}, scaling theory\cite{panyukov2018,rumyantsev2018,sayko2021}, liquid state theory \cite{perry2015,zhang2018,ylitalo2021}, transfer matrix methods\cite{lytle2017,lytle2019,chang2017,knoerdel2021}, density functional theory (DFT)\cite{wang-zhao2019}, field theoretic simulations(FTS)\cite{fredrickson2007,fredrickson2012,fredrickson2017}, and also molecular simulations\cite{hayashi2002,winkler2002,hayashi2003,hayashi2004,zhang2005,zhaoyang2006,larson2009,semenov2012,
juarez2015,peng2015,lytle2016,dzubiella2016,chang2017,radhakrishna2017,whitmer2018,
whitmer2018macro,rumyantsev2019macro,shakya2020,neitzel2021,sayko2021,bobbili2022,chen2022,chen2022-PNAS}. The emphasis strongly shifted on the presence of counterions and the interplay of entropy and enthalpy recently
\cite{perry2015,peng2015,dzubiella2016,salehi2016,fu2016,muthu2017,chang2017,radhakrishna2017,lytle2017,schlenoff2017,
meka2017,tirrell2018,adhikari2018,whitmer2018,whitmer2018macro,lytle2019,wang2019,tirrell2020,tirrell2021,mitra2023}. One may note that modeling of complex formation through bound monomer pairs considers either one-to-one monomer mapping, i.e. the `ladder' model\cite{michaels1965,winkler2002,hayashi2004,zhaoyang2006,semenov2012,dzubiella2016,chen2022,mitra2023}, or the `scrambled egg' model\cite{semenov2012,thnemann2004} of random pairing of oppositely charged monomers allowing for a higher degree of conformational freedom of the overlapping polyions, which applies progressively better for more flexible polyions\cite{semenov2012}. 
        
The debate on the thermodynamic drive for complex formation has received significant insight for strongly charged PEs, but for PEs with lower chemical charge (chemical degree of ionization or extent of ionizable monomers) content, its chemical charge dependent on the ambience, or asymmetric length, charge, or density for the pair of complexing polyions may benefit from a better understanding. Notably, some degree of nonstoichiometry is inevitable in experimental systems and a key factor to explain findings therein\cite{chen2021,ma2021}. In particular, the asymmetry may be present in the chain length or charge fractions of the complexing polyions\cite{voorn1957, juarez2015, sayko2021, chen2022,neitzel2021}, in their densities keeping the length and the charge same\cite{voorn1957,potemkin2017,zhang2018,priftis2012-softmatter,subbotin2021,friedowitz2021, bobbili2022,chen2022}, in monomer charge valency\cite{wang-zhao2019}, and so on. Early experiments\cite{kabanov1985} with complexing long polyanions as host PEs and small polycations as guest PEs found uncomplexed parts resulting from such asymmetry to be hydrophilic. Later, established theories and simulations for symmetric complexation such as the scaling theory\cite{panyukov2018,sayko2021}, random phase approximation (RPA)\cite{oskolkov2007,chen-yang2021, sayko2021}, dissipative particle dynamics (DPD) simulations\cite{juarez2015,rumyantsev2019macro, chen2022}, density functional theory (DFT)\cite{wang-zhao2019}, and liquid state (LS) theory\cite{zhang2018,ylitalo2021} have been extended with necessary modifications for non-stoichiometric coacervation, one example being the addition of one more parameter in the free energy description of the LS theory, leading to three-dimensional phase diagrams\cite{lytle2017,zhang2018}.  Simulations, on the other hand, looked into the ideal system of only two oppositely charged polyions of different sizes and variable charge densities, and gleaned important quantities like distribution of counterions, radius of gyration of polyions and the electrostatic energy gain\cite{juarez2015,chen2022}, the last one showing an increase with higher chemical charge\cite{dzubiella2016}. In addition, there are `weak' PEs with low ionizability that may be tuned with factors like solvent pH, salt, or the presence of other screening ions, leading to a variable charge ratio of the polyions\cite{lalwani2021,salehi2015,digby2022,knoerdel2021}. Recently, simulations compared the interplay of thermodynamics drives in strong and weak PEs\cite{whitmer2018,whitmer2018macro}. In addition to challenges on the energetics, soluble, two-chain complexes in asymmetric PE mixtures offer atypical morphologies such as `tadpole' complexes\cite{zhang2005,chen2022}, `fold' complexes\cite{chen2022}, and clusters of different shapes of non-zero net charge\cite{potemkin2017,panyukov2018,rumyantsev2018}, in contrast to spherical globules formed of symmetric pairs\cite{borue1988,borue1990,zhaoyang2006,chen2022,mitra2023}.

Charge stoichiometry of 1:1, leading to charge neutral complexes\cite{zhang2005}, is found to be the most favorable mixing ratio for coacervation\cite{gucht2012,priftis2012-softmatter,perry2014}. In general, asymmetric complexes stay separate when charge stabilized, and merge near charge stoichiometry, the phase separation (in the coacervate and supernatant phases) being progressively more inhibited  with increasing asymmetry\cite{voorn1957,hayashi2003,hayashi2004,chollakup2010,vitorazi2014,adhikari2018,bobbili2022,chen2022}. Simulations on a mixture of PA and PC of symmetric length and chemical charge at different concentrations\cite{hayashi2002,hayashi2003,hayashi2004,chen2022} found occurence of hierarchichally larger clusters formed due to merging of smaller clusters at closer proportions of opposite polyion charges. Two chain soluble complexes for example, which finally merge to form coacervates, will carry excess charges if there is asymmetry in chain length\cite{chen2022} or chemical charge, suggesting that the inter-complex repulsion may  make the coacervate fluidic\cite{chen2021}, leaving the 1:1 complex with the highest rigidity modulus\cite{porcel2009}. There is steep increase in coacervate densities with overall higher, symmetric charge content between the PEs\cite{neitzel2021}, indicating an increase in  binding strength with the chemical charge. In the case of small complexes formed with non-stoichiometric charge, addition of salt merges individual complexes, and at higher salt the neutral complex zone is extended\cite{zhang2005}, but if further increased salt dissolves the complex in most cases\cite{friedowitz2021,bobbili2022,chen2022}, apparently with enhanced kinetics of rearrangement\cite{kabanov1985,perry2014}. Similarly, porosity in coacervates due to nonstoichiometric charge decreases with salt, leading to elevated structural free volumes or lower volume fractions of polymers\cite{chen2021}. This has enormous impact on the mechanical properties and rigidity modulus of the assembly\cite{porcel2009}. Weaker coacervates are found to dissolve at lower salt\cite{chen-yang2021}, but for a similar degree of asymmetry in chain length compared to concentration the former is found to be more salt-resistant in inhibiting the merger of small clusters\cite{chen2022}. Furthermore, this `saloplasticity' of complex coacervates can be controlled by tuning the charge ratio of the complexing weak PEs (polyacids, for examples) by changing the pH of the solution\cite{lalwani2021,salehi2015,digby2022,knoerdel2021}, leading to solid-like PECs at high chemical charge limits\cite{lalwani2021}. The tendency to form a coacervate is found to decrease with increased charge spacing on the backbones of the PEs\cite{radhakrishna2017}.

These detailed studies reveal the richness in polyion clustering behavior in asymmetric or partially ionizable PE mixtures compared to their symmetric counterparts and motivate a more systematic investigation. It is unambiguously observed that coacervation is favored for an increase in chain length\cite{adhikari2018,gucht2012,priftis2012-softmatter,kayitmazer2015} or chemical charge density\cite{kayitmazer2015,huang2019,neitzel2021} of the complexing polyions. The size of the complex increases with asymmetry in length\cite{juarez2015} and charge density, with significant deviations from Gaussian sizes\cite{panyukov2018}. For high or low charge density in the complexing PEs one gets, respectively, solid or liquid-like PE complexes\cite{chen2021}.  Entropy of released counterions is estimated to increase with length-symmetry being restored between the PA and PC \cite{juarez2015}, but the entropic thermodynamic drive may remain inconclusive with the assumption of washed away counterions (such as in one of the RPA calculations\cite{oskolkov2007}). In other models, some counterions are found to stay within (partitioned into) the complex due to charge asymmetry\cite{shakya2020}. Furthermore, for weak Coulombic interactions which can occur in highly solvated systems, systems with well-spaced charges, or complexes in high dielectric environments, electrostatic enthalpy is predicted drive complexation\cite{zhaoyang2006}, no matter how weak the drive is, due to loosely bound counterions in these limits making the entropy gain due to the release at complexation nominal\cite{zhaoyang2006,whitmer2018,whitmer2018macro}. Requirements to identify the free energy components and to estimate the energetics of the complexation process have been argued in experiments\cite{priftis2012-softmatter}.  

In this work we aim to build a general theoretical framework to analyze the theromodynamics of the complexation of asymmetric and partially ionizable (with variable chemical charge) polyions, in the simplest case of a pair of chains. By incorporating explicit counterions and a Debye-H\"uckel treatment of ionic fluctuations, the segment-segment electrostatic as well as short-ranged excluded volume interactions, both within individual polyions and between oppositely charged polyions, are calculated leading to the Edwards' Hamiltonian\cite{edwards1979}. The ensuing, generic, free energy is subject to a variational extremization\cite{muthu1985,muthu1987,podgornik1993} by a Gaussian, trial Hamiltonian\cite{flory1950}, and it consists of contributions from the entropies of free and condensed counterions, the configurational entropy of the polyions, and the electrostatic energies of bound monomer-monomer and monomer-counterion pairs. In addition, the free energy is parameterized by the degree of asymmetry in length and number of ionizable monomers (chemical charge) of the polymers, as well as by the degree of partial ionizability (chemical charge density). First, we calculate the effective charge (defined as the net charge resulting from a part of chemical charge being partially neutralized by counterion condensation) and size of a partially ionizable single and isolated PE chain. Next, the effective charge, size, and the free energy gain of the PE complex, as a function of asymmetry in the length and then charge density of the complexing, fully ionizable chains are calculated. Further, the free energy drive of complexation as a function of the Coulomb strength of the system is explored at different charge densities, and the enthalpy- and entropy-driven regimes are identified. The crossover Coulomb strength separating the regimes is found to be marginally sensitive to the charge density. The free energy gain of complexation, calculated as a function of overlap of chains of symmetric lengths with varying, but equal, charge densities is seen to increase for lower asymmetry in the chain lengths, and a higher charge content of the complexing PEs, also observed in simulations. The theoretical framework may be useful to analyze complexations of homopolymers or intrinsically disordered proteins in dilute solutions. 	
	
\section{Model and Theory}\label{theory}
	
\begin{figure*}
	\centering
	\includegraphics[height=6.00cm,width=18.00cm]{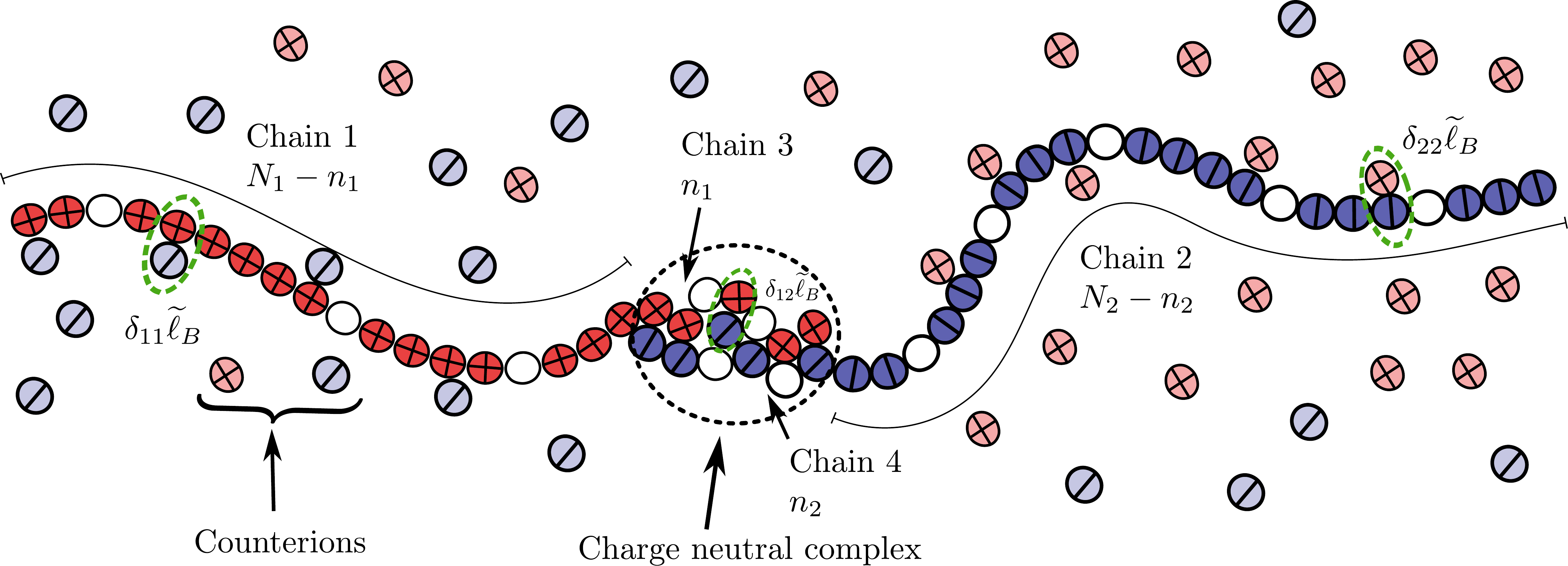}
	\caption{Schematic of partially ionizable PE chains of asymmetric length overlapping to form a complex: The oppositely charged chains slide along each other as the monomers form bound ion-pairs, releasing their counterions. $n$ is the number of ionizable monomers from both chains forming the complexed part. In the dangling chain parts, the ionizable monomers remain partially compensated by the counterions. The ion-pair formation energy, for monomer (PC)-counteranion, monomer (PA)-countercation, and monomer (PC)-monomer (PA) cases are given by $\delta_1\widetilde{\ell}_{B}$, $\delta_2\widetilde{\ell}_{B}$, and $ \delta_{12}\widetilde{\ell}_{B} $, respectively. The dangling and complexed parts have different scaling, and hence are considered to be two different chains for each polyion, resulting in the intermediate state of the complex comprising a total of four chains.}
	\label{fig:schematic}
\end{figure*}

A dilute solution of volume $\Omega$ contains two oppositely charged polymers (polyions) along with their respective counterions. In general, the polyions are asymmetric, either in size (they have different numbers of monomers, the size of the monomers being the same for both polymer chains) or in charge (chemical charge, given by the number of ionizable monomers). In addition, the polyions are in general partially ionizable (not all monomers are ionizable or charged). All charges are monovalent. Let the polyanion (PA) and polycation (PC) have $N_1, N_2$ monomers, respectively, out of which $N_{c1}, N_{c2}$ are ionizable, respectively, which are distributed uniformly along the contour of the polymer chains\cite{neitzel2021}. The complexation process of these two asymmetric and partially ionized chains proceeds as follows. Initially, the PA and PC are fully separated. The process of complexation starts with the formation of a single pair of ionizable monomers from the ends of either chain, forming one neutral monomer pair for the complex. At any general instant of time during complexation, let $n$ be the number of ionizable monomers from PA that form the intermediate complex (Fig. \ref{fig:schematic}). As the complex is taken to be electroneutral, it consists of the same number of ionizable monomers from PC. With no loss of generality let $N_{c1}<N_{c2}$. PA has $N_{c1}$ corresponding counter-cations and PC has $N_{c2}$ counter-anions, some of which may, depending on the ambient conditions (temperature, dielectricity, ion sizes etc.), be adsorbed on the respective chains, whereas the rest are free in the solution. At the same instant, $N_{c1}-n, N_{c2}-n$ ionizable monomers, respectively, (and corresponding number of total monomers - both ionizable and uncharged) remain in the dangling, uncomplexed parts of PA and PC (Fig. \ref{fig:schematic}). This overlap process continues till all $N_{c1}$ ionizable monomers from PA are exhausted and become part of the complex. In the final complex, the PC too will contribute $N_{c1}$ ionizable monomers to the complex, and there remains a dangling part of it consisting of $N_{c2}-N_{c1}$ ionizable monomers. The maximum degree of ionization are defined, respectively, as
 \begin{equation}
f_{m1}=N_{c1}/N_1,\qquad f_{m2}=N_{c2}/N_2,
\label{fmdef} 
\end{equation}
for PA and PC. This is the single-most important parameter in our work, which we may simply call the {\it ionizability} of the polyion chain, denoting the chemical charge of the polyion. The terms maximum degree of ionization, ionizability, charge density, and chemical charge will be used interchangeably in this work. 

As the ionizable monomers are randomly distributed in the respective polyions, the ratio of ionizable to uncharged monomers remains fixed for any part of the chain, whether dangling or complexed. 
In case of a fully charged homopolymer made of identical ionizable groups as repeat units, $ f_{m}=1 $\cite{mitra2023}. 

Let $M_1$ counter-cations and  $M_2$ counter-anions, respectively, remain condensed on the dangling parts of PA and PC at an intermediate state (Fig. \ref{fig:schematic}). The degree of ionization of a chain or chain part is defined as the ratio of the number of uncompensated monomer charges (charges for which the counterions are not condensed) to the number of all monomers in the chain or chain part. For PA and PC, respectively, we get the degree of ionization of the dangling parts of the chains as
\begin{equation}
f_1 = \frac{N_1 f_{m1} - M_1 - n}{N_1 - (n/f_{m1})}, \qquad f_2 = \frac{N_2 f_{m2} - M_2 - n}{N_2 - (n/f_{m2})},
\label{doidef} 
\end{equation}
where $n_i=n/f_{mi}$ is the total number of monomers from chain $i$ which form part of the intermediate complex. On the other hand, if we consider the ratio of the number of uncompensated monomer charges to the number of all monomers in the full chain, irrespective of the extent of complexation ($n_i$), we may define a total degree of ionization for the polyions as 
\begin{align}
\label{doidef_total}
		{f_{T1}}=f_{m1}-\frac{M_{1}}{N_{1}}-\frac{n}{N_{1}}, \qquad
		{f_{T2}}=f_{m2}-\frac{M_{2}}{N_{2}}-\frac{n}{N_{2}},
	\end{align}
which lead to 	
\begin{align}
\label{doidef_rel}
		f_i=\frac{N_i}{N_i-n_i}f_{Ti} \equiv \frac{N_i}{N_i-n/f_{mi}}f_{Ti}.
	\end{align}	
We may note that $f_i$ is the mean field quantity, as the bare charges on a chain are uniformly distributed only over the uncomplexed (dangling) part of the chains ($N_i-n_i$). The $n_i$ monomers in polyion $i$ are already part of the complex, and none of them can have a bare charge. Nevertheless, to get a better sense of how much charge is left in a chain after full complexation (applicable for asymmetric cases only when $N_{c1} \neq N_{c2}$), $f_{Ti}$ is a more suitable entity. In most results of this work, $f_{Ti}$ will be identified as the charge of polyion $i$. 

Further, let $c_s$ be the number density of molecules of an externally added monovalent salt that dissociates into $n_{+}$ cations and $n_{-}$ anions, where, $ c_s = n_{+}/\Omega = n_{-}/\Omega \equiv n_{s}/\Omega $. For simplicity, we assume the salt cations and anions, respectively, to be of the same type of the counter-cations and counter-anions of the PA and PC. Therefore, once salt is added,
\begin{equation}
N_{c1} - M_1 + n_s, \qquad N_{c2} - M_2 + n_s,
\label{ionsol} 
\end{equation}
cations and anions, respectively, remain free in solution. The degree of counterion condensation for the dangling chains is defined by
\begin{equation}
\alpha'_1 = \frac{M_1}{N_1-(n/f_{m1})}, \qquad \alpha'_2 = \frac{M_2}{N_2-(n/f_{m2})},
\label{docdef} 
\end{equation}
respectively, for PA and PC, which give
\begin{equation}
f_1=f_{m1}-\alpha'_1, \qquad f_2=f_{m2}-\alpha'_2.
\label{docdei} 
\end{equation}
We define two new variables related to $\alpha'_i$ which are
\begin{equation}
\alpha_1 = \frac{M_1}{f_{m1}N_1-n} = \frac{\alpha'_1}{f_{m1}}, \qquad \alpha_2 = \frac{M_2}{f_{m2}N_2-n}= \frac{\alpha'_2}{f_{m2}},
\label{docspecdef} 
\end{equation}
corresponding to the degree of condensation with respect to ionizable monomers only.
The	dimensionless parameters are defined as $\tilde{\rho}_{i}=N_{i}/(\Omega/\ell^3)$, the monomer densities of the respective chains [$(i=1,2)$] in the solution, and 
$\tilde{c}_{s}=c_s \ell^3$, where $\ell$ is the size of a monomer as well as a counterion (from either chain), and also the Kuhn length for these flexible PEs. 

Despite the complexation being a kinetic process, we make a quasistatic assumption that the conformations and degree of condensation for the dangling parts of the chains equilibrate at each step of overlap, which is parameterized by $\lambda=n/N_{c1}$. Therefore, at each step, the physical variables pertaining to the chains and the free energy components of the entire system are obtained as functions of overlap from the self-consistent minimization of the free energy with respect to polymer and counterion degrees of freedom. Relative contributions of the free energy components to complexation (i.e., difference in quantities between $n=0$, the fully separated, and $n=N_{c1}$, the maximally complexed, states) and their dependencies on electrostatic parameters (such as the Bjerrum length) are calculated. 

The intra- and inter-chain interactions, both excluded volume and electrostatic in nature, are calculated assuming that the chains follow Gaussian statistics, in terms of an expansion factor in addition to the Kuhn length. This helps us derive an approximate analytical formula for	the inter-chain pair potential in terms of the self-regulating degree of ionization and radius of gyration of the chains, and as a function of the center of mass (COM) distance between the chains (the reaction coordinate). 
	
	The total free energy $F$ of the system consisting of two PE chains, their counterions, salt ions, and the solvent (water) is assumed to have five contributions of $F_{1}$, $F_{2}, F_{3}, F_{4}$, and $F_{5}$, related, respectively, to (i) the entropy of the mobility of the adsorbed counterions along the polymer backbone, (ii) the translational entropy of the unadsorbed counterions and coions (including salt ions) which are mobile within the volume $\Omega$, (iii) the fluctuation in densities of all free small ions, (iv) the unscreened electrostatic (Coulomb) energy of the monomer-counterion and monomer-monomer bound pairs, and (v) the conformational entropy of the PEs, along with the intra-chain and inter-chain excluded volume and screened Coulomb interactions.

	%
	%
	%
%
	We consider the free energy of the system with the intermediate complex that consists of $n$ bound pairs (out of $N_{c1}$ maximum possible) and $n/f_{mi}$ total monomers from chain $i$. For the resulting dangling chain parts, the $M_{i}$ number of counterions ($i=1,2$ correspond to counter-cations to PA and counter-anions to PC, respectively) have $ W=^{f_{mi}N_{i}-n}C_{M_{i}} $ number of ways to distribute over $f_{mi}N_{i}-n$ number of ionizable monomers. The translational entropy of condensed counterions, $k_B \log W$, hence, leads to the free energy contribution
	
	\begin{align}
		\frac{F_1}{k_B T}=&\sum_{i=1}^{2}(N_{ci}-n)\left[(1-\alpha_{i})\log(1-\alpha_{i})+\alpha_i\log(\alpha_{i})\right],\label{F1_papc}
	\end{align}
where $k_B$ is the Boltzmann constant and $T$ is the absolute temperature of the system. $N_{ci}$ and $\alpha_i$ are given by Eqs. \ref{fmdef} and \ref{docspecdef}, respectively.
	
	The translational or free-volume entropy of the free ions (counterions from PA and PC and dissociated salt ions) in solution is $k_B \log[(\Omega/\ell^3 )^{\sum p_i} /\Pi p_i!] =
-k_B \Omega[ \sum c_i \log(c_i \ell^3)-\sum c_i]$, where $p_i$ is the number of ions of species (cations or anions) $i$ and $c_i \equiv p_i/\Omega$. Considering $N_{i}f_{mi}-M_{i}+n_{s}$ free ions for species $i$, the free energy contribution from free-ion entropy takes the form
 \begin{equation}
		\begin{aligned}
			&\frac{F_{2}}{k_B T}=\sum_{i=1}^{2} N_{i}\left[\left\{\left(f_{m i}-\frac{n}{N_{i}}\right)\left(1-\alpha_{i}\right)+\frac{n}{N_{i}}+\frac{\widetilde{c}_{s}}{\widetilde{\rho_{i}}}\right\}\right.\\
			&\operatorname{log}\left(\widetilde{\rho}_{i}\left(f_{m i}-\frac{n}{N_{i}}\right)\left(1-\alpha_{i}\right)+\frac{\widetilde{\rho}_{i} n}{N_{i}}+\widetilde{c}_{s}\right)\\
			&\left.-\left\{\left(f_{m i}-\frac{n}{N_{i}}\right)\left(1-\alpha_{i}\right)+\frac{n}{N_{i}}+\frac{\widetilde{c}_{s}}{\widetilde{\rho_{i}}}\right\}\right].
		\end{aligned}\label{F2_papc}
	\end{equation}
	
The free energy contribution due to the fluctuations in
the densities of the same set of free ions ($N_{i}f_{mi}-M_{i}+n_{s}$) is given by the Debye-H\"uckel expression\cite{mcquarrie2000},
	\begin{align}
		&\frac{F_{3}}{k_{B} T}=-\frac{\Omega\kappa^3}{12\pi}
		=\frac{-2\sqrt{\pi}\widetilde{\ell}_{B}^{3/2}N_{1}}{3\widetilde{\rho}_{1}}  \nonumber\\
		& \times \left[\sum_{i=1}^{2}\left(\widetilde{\rho}_{i}\left(f_{m i}-\frac{n}{N_{i}}\right)\left(1-\alpha_{i}\right)+\frac{\widetilde{\rho}_{i} n}{N_{i}}+\widetilde{c}_{s}\right)\right]^{3/2},\label{F3_papc}
	\end{align}	
where $\tilde{\ell}_B=\ell_B/\ell$ is the dimensionless Bjerrum length (or the `electrostatic temperature') where $\ell_B=e^2/4 \pi \epsilon_0 \epsilon k_B T$ is the Bjerrum length with $e$ the electronic charge, $\epsilon_0$ the dielectric permittivity of free space, $\epsilon$ the dielectric constant, and $\ell$ the Kuhn length. The inverse Debye length $\kappa$ is defined as
	\begin{align}\label{kappa}
		\tilde{\kappa}^{2}=&4 \pi  \tilde{\ell}_{B} \sum_p z_{p}^{2} n_{p} / \Omega\nonumber\\
		=&4 \pi \tilde{\ell}_{B} \left[\sum_{i=1}^{2}\widetilde{\rho}_{i}\left(f_{m i}-\frac{n}{N_{i}}\right)\left(1-\alpha_{i}\right)+\frac{\widetilde{\rho}_{i} n}{N_{i}}+\widetilde{c}_{s}\right].
	\end{align}
	
Here, $\tilde{\rho}_{i}=N_{i}/(\Omega/\ell^3)$, $\widetilde{\kappa}=\kappa \ell$, $n_p$ is the number, and $ z_{p}$ is the valency of the dissociated ions of the $p^{\text{th}}$ species. 
	


	
	The energy of the three types of ion-pairs - negatively charged monomer and counter-cation, positively charged monomer and counter-anion, and oppositely charged monomers (Fig. \ref{fig:schematic})- are of the Coulombic form $-e^2/4 \pi \epsilon_0 \epsilon_\ell d$, where $d$ is the dipole length of the pair and $\epsilon_\ell$ the local dielectric constant in the vicinity of the ion-pairs.
Summing over all ion-pairs of three types for the complex and the dangling chains, the electrostatic free energy of the ion-pairs is obtained as
	\begin{align}
		\frac{F_{4}}{k_{B} T}&=-\sum_{i}N_{i}\left\{\left(f_{mi}-\frac{n}{N_{i}}\right)-\left(1-\frac{n_{i}}{N_{i}}\right)f_i \right\}\widetilde{\ell}_{B}\delta_{i}\nonumber\\
		&- n \widetilde{\ell}_{B} \delta_{12}, \label{F4_papc}
	\end{align}
where $\delta_i$'s are the dielectric `mismatch' parameters for the monomer-counterion pairing
in the dangling chains, and $\delta_{12}$ is the same parameter for the monomer-monomer pairing
in the complexed part. Note that the two terms in the sum are actually $-M_i\widetilde{\ell}_{B}\delta_{i}$. The organic backbone of the polyions, which are isotropic fractals, disturbs in its vicinity the
orientational ordering of solvent dipoles in polar solvents, which results in a local
dielectric constant that is lower compared to the bulk\cite{khokhlov1994,kramarenko2002,muthu2004,delacruz2004,arindam2010}. In general, 
the $\delta=(\epsilon/\epsilon_\ell)(\ell/d)$ represents both the disparity between the local ($\epsilon_\ell$) and bulk ($\epsilon$) dielectric constants of the medium and the measure of the ion sizes (contributing to the dipole length $d$) that strongly influences the interaction strength of ion pairs\cite{mitra2023}. $\delta$, in general, is different for different solvent-PE pairs as well as for different PE-counterion pairs. The bound pair energy in $F_4$ then turns out to be proportional to the product $\delta \tilde{\ell}_B$ which is considered as the `electrostatic' or `Coulomb' strength of the system in this work. The majority of this work assumes a constant $\delta$, and the electrostatic strength will mainly be interpreted by the Bjerrum length $\ell_B$. The Coulomb strength sensitively controls the screening effect due to salt, but does not depend on the salt concentration (which sets the ionic strength of the solution). 

In addition to the free energy components $F_1$ to $F_4$ related to the counterion degrees of freedom, there is conformational entropy of the individual polyion chains and the interaction energy among the two chains related to the polymer degrees of freedom. As the size scaling and monomer distributions of the dangling and complexed parts, in each polyion (Fig. \ref{fig:schematic}), are expected to be different in general, we consider them to be two different chains within the same PE to calculate the interactions, and to apply the Gaussian trial Hamiltonian (as we shall see later). Therefore, in principle, we have four polymer chains (given by indices 1 to 4), the two of which (1 and 2) are the two dangling parts, respectively, and the other two (3 and 4) form the complexed part, of the two original polyions. We expect the free energy $ F_{5} $, originating from the Hamiltonian at a particular degree of overlap (at which $n_i$ molecules from chain $i$ form the complexed part of the chains), to be a function of the charges ($f_{i}$) of the dangling parts of the polyions and the configurational distribution of monomers of all `four' chains (in terms of the size expansion factors $\tilde{\ell}_{1i}$, $i$=1 to 4, to be defined later). It also must comprise parameters such as chain lengths ($N_i$), salt concentration ($\tilde{c}_s$), Bjerrum length ($\tilde{\ell}_B$), and those related to excluded volume interactions. The overall free energy ($F_5$) must also depend on the reaction coordinate (COM distance between the chains).
			%
			
			The Hamiltonian $H$ can be written in terms of the following monomer-monomer interactions - a) their connectivity ($ H_{0} $), b) repulsive, non-electrostatic excluded volume interactions among all monomers ($ H_{ex} $) and c) electrostatic interactions between charge uncompensated monomers ($ H_{el} $) - repulsive for like-charged monomers within a chain (intra-chain - either PA or PC), but attractive for oppositely charged monomers from complexing chains (inter-chain - both PA and PC). Hence 
			\begin{align}\label{eq:H}
				&{\beta H}=H_{0}+H_{ex}+H_{el},
			\end{align}
			where the components ($H_{0},~H_{ex}\text{ and }H_{el}$) are given by
			\begin{widetext}
				\begin{align}
					&H_{0}=\sum_{i=1}^{4}\frac{3}{2 \ell^2}\int_{0}^{\mathcal{N}_{i}} d s_{i}\left(\frac{\partial {\bf R}\left(s_{i}\right)}{\partial s_{i}}\right)^{2},\label{eq:H_0}\\
					&H_{ex}=\sum_{i=1}^{4}\omega_{ii}\ell^{3} \int_{0}^{\mathcal{N}_{i}} ds_{i} \int_{0}^{\mathcal{N}_{i}} ds_{i}^\prime \delta( \mathbf{R}(s_{i})-\mathbf{R}(s_{i}^\prime))+2\omega_{34}\ell^{3} \int_{0}^{\mathcal{N}_{3}} ds_{3} \int_{0}^{\mathcal{N}_{4}} ds_{4} \delta(\mathbf{R}(s_{3})-\mathbf{R}(s_{4})),\label{eq:H_ex}\text{ and }\\
					&H_{el}=\frac{1}{2}\sum_{i=1}^{2}\int_{0}^{\mathcal{N}_{i}} ds_{i} \int_{0}^{\mathcal{N}_{i}} ds_{i}^\prime f_{i}f_{i}\ell_{B}\frac{ e^{-\kappa \left|{\bf R}(s_i)-{\bf R}(s_{i}^\prime)\right|}}{\left|{\bf R}(s_i)-{\bf R}(s_{i}^\prime)\right|} + \int_{0}^{\mathcal{N}_{1}} ds_{1} \int_{0}^{\mathcal{N}_{2}} ds_{2}   f_{1}f_{2}\ell_{B}\frac{ e^{-\kappa \left|{\bf R}(s_1)-{\bf R}(s_2)\right|}}{\left|{\bf R}(s_1)-{\bf R}(s_2)\right|}.\label{eq:H_el}
				\end{align}
			\end{widetext}
$\mathcal{N}_i$ are the number of monomers in chains $i$=1 to 4. $\mathcal{N}_i=N_i-n_i$
for the dangling parts (chains $i=1,2$), $\mathcal{N}_3=n_1$, and $\mathcal{N}_4=n_2$, where $n_i=0$ represent the uncomplexed polyions PA and PC. The arguments of the $\delta$-functions in the above integrals are the difference in contour vectors corresponding to the monomer pair involved in the excluded volume interaction. The charge uncompensated monomers interact through a screened Coulomb electrostatic potential. The excluded volume contribution among monomer-pairs within individual PE molecules is given by the first four terms in the r.h.s of Eq. \ref{eq:H_ex}, corresponding to interaction strengths $ w_{11},w_{22},w_{33}$, and $ w_{44} $, and will sensitively depend on the configurations of individual chains. The mutual excluded volume interaction for the polyion pair is given by the fifth term, corresponding to the interaction strength $ w_{34} $. The other contributions corresponding to $w_{12}, w_{13}, w_{14}, w_{23}$, and $w_{24}$ are ignored as the polyions in those pairs of chains (in our four-chain model) are separated from each other without any overlapping monomers (Fig. \ref{fig:schematic}).  Chains 3 and 4 are considered to be uncharged, as the charged monomers therein are assumed to have formed ion-pairs as part of the complex. As mentioned before, the complexed part made of monomer-monomer ion-pairs is assumed to be charge neutral ($ f_{m1}n_{1}=f_{m2}n_{2} $) and not to affect the electrostatic interactions between the uncomplexed parts of the polyions. We note that the monopole-dipole or dipole-dipole interactions are short-ranged similar to excluded-volume interaction \cite{pincus1998,winkler1998,liu2002,muthu2004,kundu2014,mitra2023}. As globular collapse of PE chains in the complex is known not to affect the thermodynamics significantly\cite{mitra2023}, the collapsed state will not be considered explicitly, and the dipolar contributions will be ignored compared to the monopole-monopole interaction in this work.  In Eq. \ref{eq:H_el}, the pair of charged monomers may come from the same chain (the first two terms) or from two different chains (the third term). The counterion condensation on the uncomplexed monomers has been addressed at the mean field level, leading to the coefficients of the general type $f_if_j$, for the two terms in Eq. \ref{eq:H_el}.

It is rather tedious to directly evaluate the partition sum using the Hamiltonian discussed above (Eqs. \ref{eq:H}, \ref{eq:H_0}, \ref{eq:H_ex}, \ref{eq:H_el}). Instead, following the variational procedure\cite{muthu1987} one may introduce a trial Hamiltonian\cite{muthu1987} by rewriting the Hamiltonian given in Eq. \ref{eq:H} as 
			\begin{align}
				H=&H_{trial}+(H-H_{trial})
			\end{align}
where
			\begin{align}\label{trial-Hamiltonian}
			H_{trial}=\sum_{i=1}^{4}\frac{3}{2 \ell\ell_{1i}}\int_{0}^{\mathcal{N}_{i}} d s_{i}\left(\frac{\partial {\bf R}\left(s_{i}\right)}{\partial s_{i}}\right)^{2}.
			\end{align}
$\ell_{1i}$ is the variational parameter that gives the effective expansion factor of the polyion compared to its Gaussian size\cite{edwards1979,muthu1987,muthu2004}. 
			The mean-field assumption is based on the (Gibbs-Bogoliubov) inequality,
			\begin{align}
				\left\langle\mathrm{e}^{-\beta H}\right\rangle_{H_{trial}} \geq \mathrm{e}^{-\beta\left\langle H\right\rangle_{H_{trial}}},
			\end{align}
which implies that the extremum free energy,
\begin{align} \label{F5-trial}
\widetilde{F}_5=\langle \beta(H_0-H_{trial})\rangle_{H_{trial}}+\langle \beta H_{ex}\rangle_{H_{trial}}
+\langle \beta H_{el}\rangle_{H_{trial}},
\end{align}
needs to be extremized with respect to the charges of the two original polyions ($ f_1,f_2 $) and sizes of the four chains ($ \ell_{11},\ell_{12},\ell_{13},\ell_{14} $) (see Fig. \ref{fig:schematic}). If one shifts to the polymer coordinate (the spatial coordinate ${\bf r}$) the Hamiltonian can be approximately recast in terms of the monomer density profiles of the four chains\cite{podgornik1993}. Taken as a spherically-symmetric Gaussian distribution, the monomer density, centered at $ \mathbf{r}_{i}$, at a position $ \mathbf{r} $ is given by
			\begin{align}\label{rho-def}
				&\rho_{ni}(\ro)=\mathcal{N}_{i}\left(\frac{3}{4 \pi R_{gi}^{2}}\right)^{3 / 2} \exp \left[-\frac{3 (\left|\ro-\ro_{i}\right|)^{2}}{2 R_{gi}^{2}}\right]
			\end{align}
			 ($i=1$ to 4), where the number of monomers in the $i$-th chain is $ \mathcal{N}_{i} $. Within the approximation of uniform expansion of the PE chains, the average dimensionless radii of gyration of the chains are given by
			\begin{align}\label{rg_def}
				\widetilde{R}_{gi}=\sqrt{\frac{\mathcal{N}_{i}  \tilde{\ell}_{1 i}}{6}},
			\end{align}
where $\tilde{\ell}_{1 i}=\ell_{1i}/\ell$.			
			
Using the Fourier transforms (in ${\bf k}$-space) of the monomer density profiles (Eq. \ref{rho-def}) and integrating the averaged interaction of monomers (Eq. \ref{F5-trial}), the total free energy contribution due to the polymer degrees of freedom included in the Hamiltonian (Eq. \ref{eq:H}) can be arrived in the form 
				\begin{align}\label{eq:F5}
					&\beta F_{5}=\frac{3}{2}\sum_{i=1}^{4}\left(\widetilde{\ell}_{1i}-1-\log \widetilde{\ell}_{1i}\right)+\left(\frac{9}{2\pi}\right)^{3/2} \sum_{i=1}^{4} \frac{w_{ii}{\mathcal{N}_{i}}^{1/2}}{\widetilde{\ell}_{1i}^{3/2}}\nonumber \\
					&+w_{34}\mathcal{N}_3 \mathcal{N}_4 \left(\frac{3}{4\pi \widetilde{R}_{g0}^2}\right)^{3/2}\exp\left(-\frac{3\widetilde{R}_{34}^2}{4 \widetilde{R}_{g0}^2}\right)\nonumber \\
					&+\sum_{i=1}^{2}\frac{{f_{i}}^{2} \mathcal{N}_{i}^{2}\widetilde{\ell}_{B}}{2} \Theta_{s}\left(\widetilde{\kappa},a_{i}\right)- {f_{1}} {f_{2}} \mathcal{N}_{1}\mathcal{N}_{2}\widetilde{\ell}_{B}\Theta_{m}\left(\widetilde{\kappa},\widetilde{R}_{12},a_{12}\right),
				\end{align}
where, $i$=1 to 4 correspond to the four-chain description (Fig. \ref{fig:schematic}).  The first summed term corresponds to the conformational entropy of the chains ($F_{51}/k_B T$), the $w_{ij}$ terms to the excluded volume interaction among all monomers ($F_{52}/k_B T$), and the $f_i f_j$ terms to screened electrostatic interaction among charge-uncompensated monomers within individual polyions and between polyions ($F_{53}/k_B T$).

In Eq. \ref{eq:F5}, $ \widetilde{R}_{12}={\left|\ro_{1}-\ro_{2}\right|}/{\ell}\equiv \widetilde{R}_{g1}+\widetilde{R}_{g2}+2\widetilde{R}_{g4}$ is the dimensionless separation distance between the two dangling parts of the polyions (see Eq. \ref{rg_def} and Fig. \ref{fig:schematic}), where the dimensionless radius of gyration $ \widetilde{R}_{gi} $ is defined by Eq. \ref{rg_def}. Further, $ \widetilde{R}_{g0}^{2}=({n_{1}\widetilde{\ell}_{1 3}+n_{2}\widetilde{\ell}_{1 4}})/{6}$ (in Eq. \ref{eq:F5}) and $ \widetilde{R}_{34}=\left|{\bf r}_{3}-{\bf r}_{4}\right|/\ell \equiv 0 $. $\Theta_{s}\left(\widetilde{\kappa},a_{i}\right)$ is given by  
			\begin{equation}\label{Theta_s}
				\Theta_{s}\left(\widetilde{\kappa},a_{i}\right)=\frac{2}{\pi}\left[\sqrt{\frac{\pi\widetilde{\kappa}^{2}}{4 a_{i}}}-\frac{\widetilde{\kappa} \pi}{2} \exp{\left(a_{i}\right)} \text{erfc}\left(\sqrt{a_{i}}\right)\right],
			\end{equation}
			and $\Theta_{m}\left(\widetilde{\kappa}, \widetilde{R}_{12},a_{12}\right)$ by
			\begin{align}\label{Theta_m}
				\Theta_{m}\left(\widetilde{\kappa}, \widetilde{R}_{12},a_{12}\right)=&\frac{e^{a_{12}}}{ \widetilde{R}_{12}}\left[e^{-\widetilde{\kappa}  \widetilde{R}_{12}} \text{erfc}\left(\sqrt{a_{12}}-\frac{\widetilde{\kappa}  \widetilde{R}_{12}}{2 \sqrt{a_{12}}}\right)\right.\\ \nonumber
				&\left.-e^{\widetilde{\kappa}\widetilde{R}_{12}} \text{erfc}\left(\sqrt{a_{12}}+\frac{\widetilde{\kappa}  \widetilde{R}_{12}}{2 \sqrt{a_{12}}}\right)\right],
			\end{align}
where the quantities $a_{i}$ and $a_{12}$ are given by 
			\begin{align}
				&a_{i}=\widetilde{\kappa}^2 \widetilde{R}_{gi}^{2}/3=\widetilde{\kappa}^{2}\widetilde{\ell}_{1i}(N_{i}-n_{i})/18, \text{\; and}\\
				&a_{12}=\sum_{i=1}^{2}\widetilde{\kappa}^{2}\widetilde{R}_{gi}^{2}/6=\sum_{i=1}^{2}\widetilde{\kappa}^{2}\widetilde{\ell}_{1i}(N_{i}-n_{i})/36.
			\end{align} 

As constructed, the free energy is a function of the variables $\widetilde{\ell}_{11}$, $\widetilde{\ell}_{12}$, $\widetilde{\ell}_{13}$, and $\widetilde{\ell}_{14}$, which are the effective expansion factors compared to the Gaussian size for the mean square end-to-end distance of the dangling and complexed parts of the chains (chains 1 to 4, Fig. \ref{fig:schematic}), and also a function of the degree of ionization $f_1$ and $f_2$ of the dangling parts, respectively.  
			
Finally, we may write the total free energy of the system consisting of the two complexing polyions, its counterions, salt ions, and the solvent as 
			\begin{align}
				\frac{F}{k_{B}T}=\sum_{i=1}^{5} \frac{F_{i}}{k_{B}T},\label{Ftotal}
			\end{align}
			where $F_1, F_2, F_3, F4$ and $F_5$ are given by, respectively, Eqs. \ref{F1_papc},\ref{F2_papc},\ref{F3_papc},\ref{F4_papc} and \ref{eq:F5}.

\section{Results and discussion }	
	

The equilibrium conformations (and hence the effective expansion factors) of the oppositely charged polyions and their effective charge, regulated by the degree of counterion condensation,  are determined in this two-state model (the counterions are either bound on the chain backbone or free in the solution), by the self-consistent minimization of the free energy (Eq. \ref{Ftotal}) ideally with respect to six variables - the charge ($ f_{Ti} $) and size ($\widetilde{\ell}_{1i}$) of the PEs, which are divided into the dangling and complexed parts at an intermediate step of complexation (Fig. \ref{fig:schematic}). As the chains in the complexed part (chains 3 and 4 in Fig. \ref{fig:schematic}) are assumed to take Gaussian conformations, if excluded volume interactions are ignored considering the system to be mostly driven by charge interactions (actually they are sub-Gaussian chains or compact globules depending on the charge-correlations, such as dipolar interactions, that we ignore in the context of the energetics\cite{mitra2023}), $\widetilde{\ell}_{13}$ and $\widetilde{\ell}_{14}$ can be taken as 1, which leads to a four variable minimization with respect to $ f_{T1}, f_{T2}, \widetilde{\ell}_{11}, \widetilde{\ell}_{12}$.  The minimization yields the equilibrium size and charge of the polyions - PA and PC - at different overlap (specified by the parameter $\lambda=n/N_{c1}$), including the completely separated ($\lambda=0$) and fully complexed ($\lambda=1$) chains. For all the results presented in the paper, the dielectric mismatch parameter is chosen to be the same for the all ion-pair combinations - PA-counter-cation, PC-counter-anion and PA-PC monomers - leading to $\delta_1=\delta_2=\delta_{12} \equiv \delta$.
The excluded volume parameters for all interactive pairs - monomer pair within PA, monomer pair within PC, and monomer pair with one monomer from each PA and PC - have been taken to be zero, considering that the electrostatic interactions will continue to overwhelm excluded volume interactions,
especially at good solvents for most conditions\cite{mitra2023}. Hence, $w_{11}=w_{22}=w_{12}=0.0$ or
$w_{ij}=0.0$ are assumed for the rest of the paper. To estimate the monomer densities of the individual chains, we set the dimensionless volume of the system to be $\Omega/\ell^3=2\mathrm{x}10^6$. For the $i$-th polyion of length $N_i$ (here, $i=1,2$ corresponding to the full PA and PC), the monomer density is given by $\tilde{\rho}_i =N_i/(\Omega/\ell^3)$. For $N_i=1000$, it turns out to be $\sim 0.0005$.

	\subsection{Single polyelectrolyte chain: effect of ionizability $f_{m}$}

Before analysing the thermodynamics of complexation of two partially ionizable, oppositely charged PE chains, we look into the ionization equilibrium for a single such polyion in isolation. The total free energy of the system consisting of an isolated and flexible PE chain with ionizable monomers, counterions, and the added salt ions (all ions are monovalent) in a solvent at high dilution can be written as $F=\Sigma_{i} F_{i}$\cite{arindam2010}, where,
\begin{align}
	\frac{F_{1}}{N k_{\mathrm{B}} T}&=\left(f_{m}-\alpha\right) \log \left(1-\frac{\alpha}{f_{m}}\right)+\alpha \log \left(\frac{\alpha}{f_{m}}\right),
\end{align}
\begin{align}
	\frac{F_{2}}{Nk_{\mathrm{B}}T}&=\left(f_{m}-\alpha+\frac{\tilde{c}_{s}}{\tilde{\rho}}\right) \log\left\{\tilde{\rho}\left(f_{m}-\alpha\right)+\tilde{c}_{s}\right\} \nonumber\\
	&+\frac{\tilde{c}_{s}}{\tilde{\rho}} \log \tilde{c}_{s}-\left(f_{m}-\alpha+2 \frac{\tilde{c}_{s}}{\tilde{\rho}}\right),
\end{align}
\begin{align}
	\frac{F_{3}}{N k_{\mathrm{B}} T}&=-\frac{2}{3} \sqrt{\pi}\frac{\tilde{\ell}^{3 / 2}_{B}}{\tilde{\rho}}\left\{\tilde{\rho}\left(f_{m}-\alpha\right)+2 \tilde{c}_{s}\right\}^{3 / 2},
\end{align}
\begin{align}
	\frac{F_{4}}{N k_{\mathrm{B}} T}&=-\alpha \tilde{\ell}_{B} \delta,\text{ and}
\end{align}
\begin{align}
	\frac{F_{5}}{N k_{\mathrm{B}} T}&=\frac{3}{2 N}\left(\tilde{\ell}_{1}-1-\log \tilde{\ell}_{1}\right) +\left(\frac{9}{2 \pi}\right)^{3 / 2} \frac{w}{ \sqrt{N} \tilde{\ell}^{3 / 2}_{1}}\nonumber\\
	&+\frac{({f_{m}}-\alpha)^{2} \widetilde{\ell}_{B}N}{2} \Theta_{s}\left(\widetilde{\kappa}, a\right), \nonumber\\
	\text { where } &\Theta_{s}\left(\widetilde{\kappa},a\right)=\frac{2}{\pi}\left[\sqrt{\frac{\pi\widetilde{\kappa}^{2}}{4 a}}-\frac{\widetilde{\kappa} \pi}{2} \exp{\left(a\right)} \text{erfc}\left(\sqrt{a}\right)\right]
	\label{F5_single}
\end{align}
with the contributions being due to, respectively, the entropy of mobility along the chain backbone of adsorbed counterions $\left(F_{1}\right)$, the translational entropy of the mobile ions ($F_{2}$, including both free counterions from the polymer and the salt ions), fluctuation in densities of all mobile ions $\left(F_{3}\right)$\cite{mcquarrie2000},  adsorption (Coulomb) energy of the bound counterion-monomer pairs $\left(F_{4}\right)$, and the free energy of the chain $\left(F_{5}\right)$ resulting from its connectivity (conformational entropy) (first term in $F_{5}$), the two-body excluded volume interaction (second term in $F_{5}$), and the screened electrostatic interactions between charge-uncompensated monomers ($N-M$) (third term in $F_{5}$).
$N,N_c,M$ are numbers, respectively, of monomers, ionizable monomers, and condensed counterions. Here, $f_{m} \equiv N_{c} / N$ is the ionizability, $\alpha \equiv M/N$ the degree of condensation, and $f=(N_c-M)/N \equiv f_m - \alpha$ the degree of ionization (`charge'), $\tilde{\ell}_1=6 R_g^2/N\ell^2$ the expansion factor of the polyion chain (`size'), and $\tilde{\rho} \equiv \rho\ell^{3} \equiv N \ell^{3} / \Omega$ the monomer concentration, where $R_g$ is the radius of gyration of the chain. $\delta=\left(\varepsilon \ell / \varepsilon_{\ell} d\right)$, the dielectric mismatch parameter, and $\Theta_{s}\left(\widetilde{\kappa},a\right)$, given by Eq. \ref{Theta_s}, have usual meaning as defined before (in section \ref{theory}). 
In addition, $a=\tilde{\kappa}^2 N \tilde{\ell}_1/18$, where
\begin{equation}
	\tilde{\kappa}=\sqrt{4 \pi \tilde{\ell}_{B}\left\{\tilde{\rho}\left(f_{m}-\alpha\right)+2 \tilde{c}_{s}\right\}},
\end{equation}
is the dimensionless inverse Debye screening length for this single chain system (see Eq. \ref{kappa} for the two-chain system), where $\tilde{\kappa}=\kappa \ell$.  

The charge ($f$) and size ($\tilde{\ell}_1$) of a single, isolated, flexible, and fully ionizable ($f_m=1$) polyelectrolyte chain in solution are well-understood quantities in PE theory\cite{muthu2004, arindam2010}. One typical way to derive such quantities is to apply a self-consistent double minimization of the free energy ($F$) with respect to charge and size. However, in the adiabatic approximation\cite{arindam2010}, applicable to the expanded chain conformations in good solvents, only $F_1,F_2$ and $F_4$ will contribute to the minimization of the charge ($f$) independently of size ($\tilde{\ell}_1$), and one shall have the analytical expression of charge
\begin{eqnarray}
\label{single-chain-charge-anal}
f=\frac{-\left(\tilde{c}_s+e^{-\delta \tilde{\ell}_B}\right)
+ \sqrt{\left(\tilde{c}_s+e^{-\delta \tilde{\ell}_B}\right)^2+ 4 \tilde{\rho} f_m e^{-\delta \tilde{\ell}_B}}}{2 \tilde{\rho}},
\end{eqnarray}
which can be plugged back in $F_5$ in Eq. \ref{F5_single} to determine the size ($\tilde{\ell}_1$) of the polyion.

\begin{figure}[!htbp]%
	\centering%
	\subfloat[]{\includegraphics[height=4.00cm,width=4.25cm]{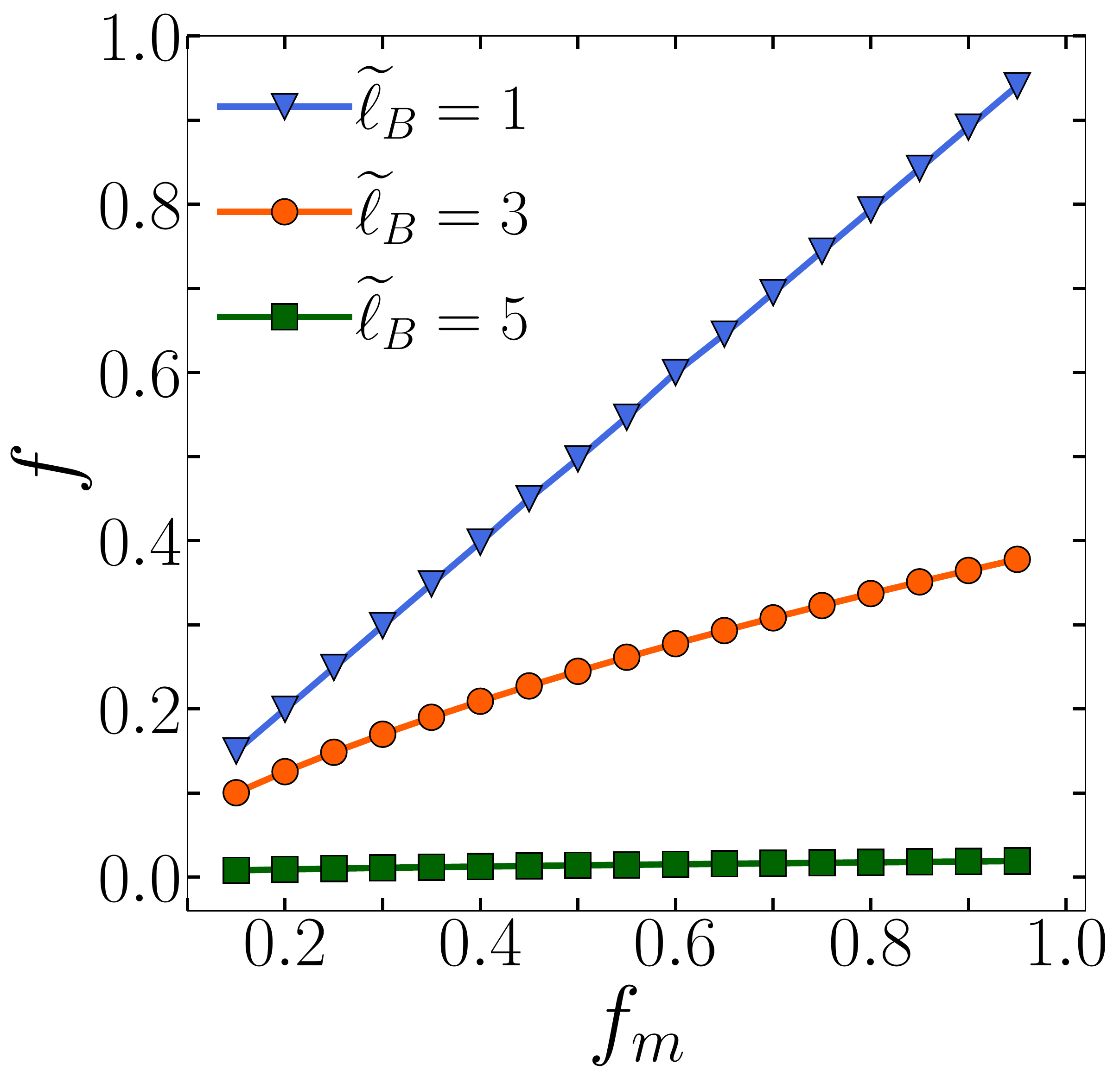}}%
	\subfloat[]{\includegraphics[height=4.00cm,width=4.25cm]{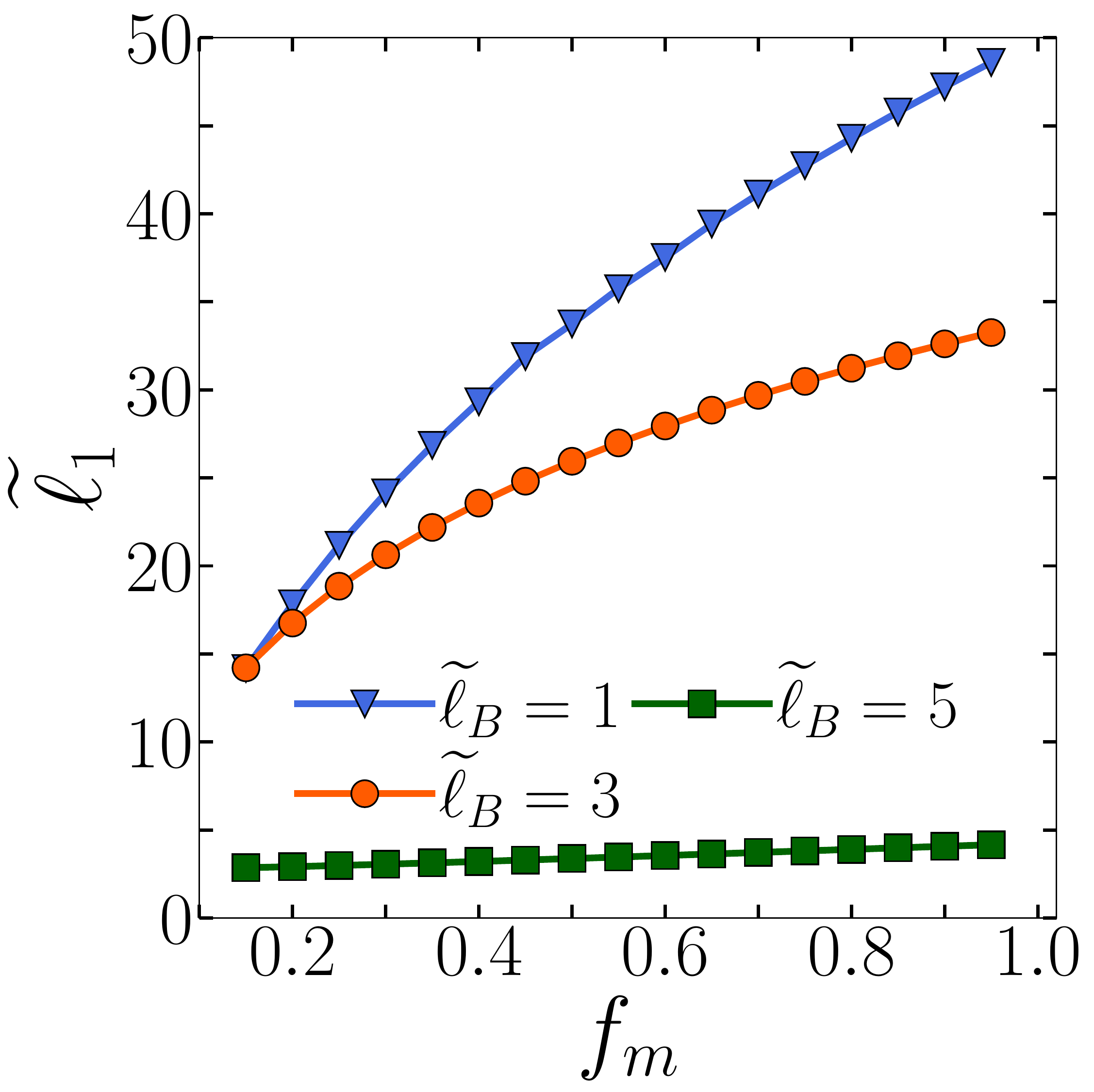}}%
	\caption{Partially ionizable single polyelectrolyte: The charge ($f$) and size ($\widetilde{\ell}_{1}$) of the PE chain are plotted as functions of ionizability $f_{m}$, for three values of the
	Coulomb strength $\tilde{\ell}_B=1,3,5$. For this range of $\tilde{\ell}_B$, both size and charge increase with decreasing $\tilde{\ell}_B$ and increasing ionizability. The other parameters are: $N=1000, N_c=1000, \delta=3.5, w=0, \tilde{c}_s=0$, and $\tilde{\rho}=0.0005$.}%
	\label{fig:scpar}%
\end{figure}

The only difference from the previous results of fully ionizable chains\cite{muthu2004,arindam2010} we have is that, unlike fully charged PEs, the PE chain can be partially ionizable, in principle with a very low effective charge density. The charge and size, for fixed Coulomb strengths ($\tilde{\ell}_B=1,3,5$, $\delta=3.5$) and zero salt, are plotted as a function of ionizability ($f_m$) in Fig. \ref{fig:scpar}. As the number of free counterions and uncompensated monomers increase monotonically with $f_m$ for constant Coulomb strengths, so does $f$ and consequently $\tilde{\ell}_1$ of the PE molecule (the analytical expression of $f$, Eq. \ref{single-chain-charge-anal}, closely follows the full numerical results). For a given value of ionizability, at a lower $\widetilde{\ell}_{B}=1$ (higher temperatures), the counterions are mostly free in solution enjoying volume entropy, and very few of them are bound to the chain monomers, which results in a higher charge and size. For high Coulomb strengths the electrostatic effects are suppressed due to counterion condensation, and the profiles become insignificant like uncharged chains. The range of $\widetilde{\ell}_{B}$ used here does not include very low values, for which the monomers are completely thermalized, and the PE chain takes the Gaussian size of maximum entropy even if highly charged. The values of the charge and size of a single chain obtained here will be more or less applicable to individual polyions participating in two-chain complexation, when they are separated, just before the onset of the process. 

The parameters chosen here such as the Bjerrum length $\widetilde{\ell}_{B}= 3$, the middle value, and $\delta=3.5$, correspond closely to the case of a strong, flexible PE, sodium(polystyrene sulfonate), NaPSS, in aqueous solutions at room temperature\cite{muthu2004}. 

	\subsection{Fully ionizable polyelectrolyte chains and complexation}
	
	\subsubsection{Symmetric polyelectrolyte chains: equal length}
	
	We first benchmark the general problem of complexation of two oppositely charged, partially ionizable PEs of different length by the known results for fully ionizable and symmetric PEs (of same length and number of monomers).  The correlation between counterion adsorption, release and pairwise electrostatic attraction at various ambient conditions has already been investigated in details\cite{mitra2023}. Here, we briefly note the key results, obtained by the minimization of the free energy (Eq. \ref{Ftotal}) for the symmetric, fully ionizable case (the ionizability $f_{m1}=f_{m2}=1$, $f_{T1}=f_{T2}=f_T$, and $\tilde{\ell}_{11}=\tilde{\ell}_{12}=\tilde{\ell}_{1}$), rendering the minimization to be effectively with respect to two variables $f_T$ and $\tilde{\ell}_{1}$. The polyions have the same number of monomers $N_1=N_2=N$, and are fully ionizable implying $N_{c1}=N_{c2}=N$. As mentioned before, 
	$\delta_1=\delta_2=\delta_{12} \equiv \delta$ and $w_{11}=w_{22}=w_{12}=0.0$ or
$w_{ij}=0.0$. In addition, for the symmetric chains of equal length, $\tilde{\rho}_1=\tilde{\rho}_2 \equiv \tilde{\rho}$. 

\begin{figure}[!htbp]%
	\centering%
	\subfloat[]{\includegraphics[height=3.50cm,width=4.25cm]{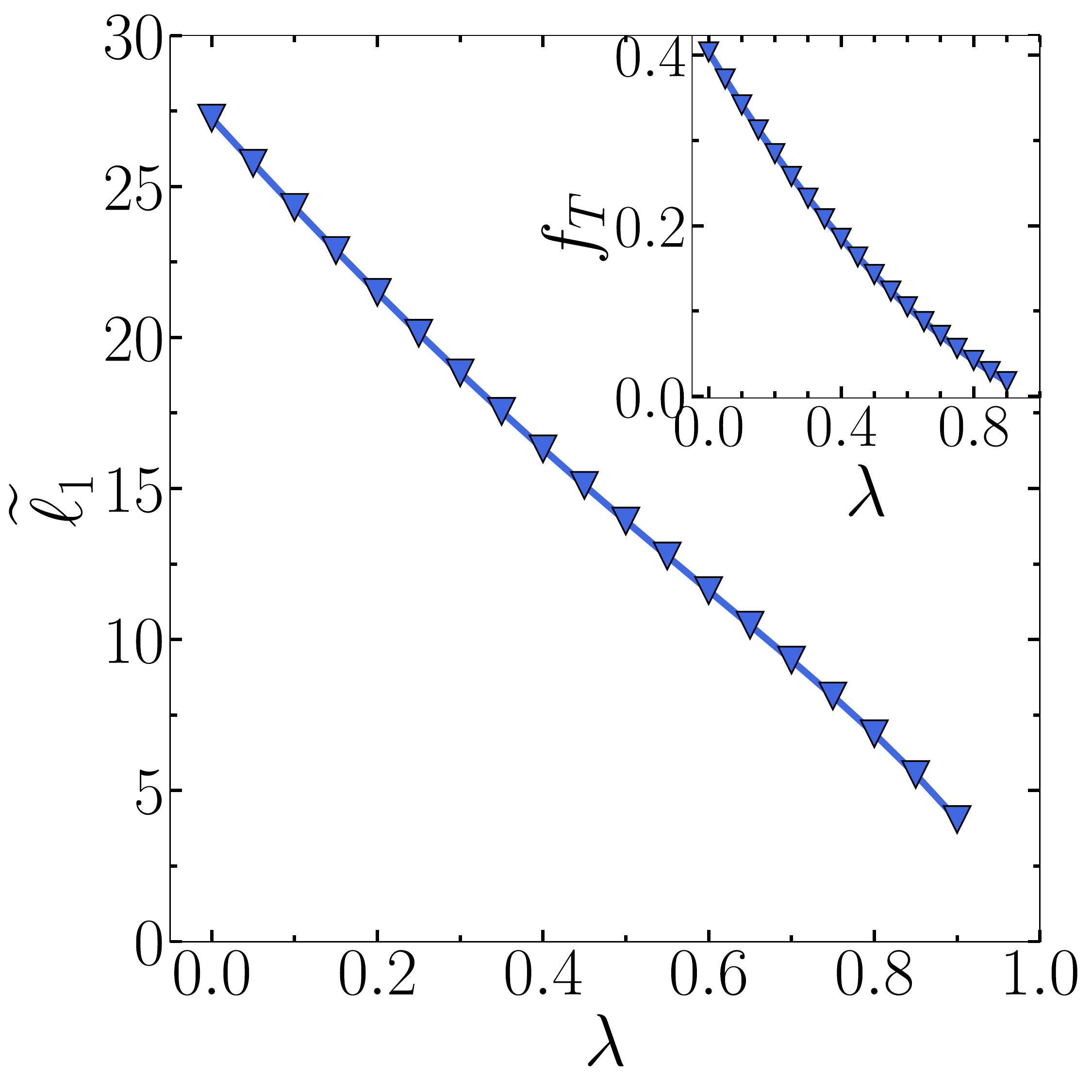}}%
	\subfloat[]{\includegraphics[height=3.50cm,width=4.25cm]{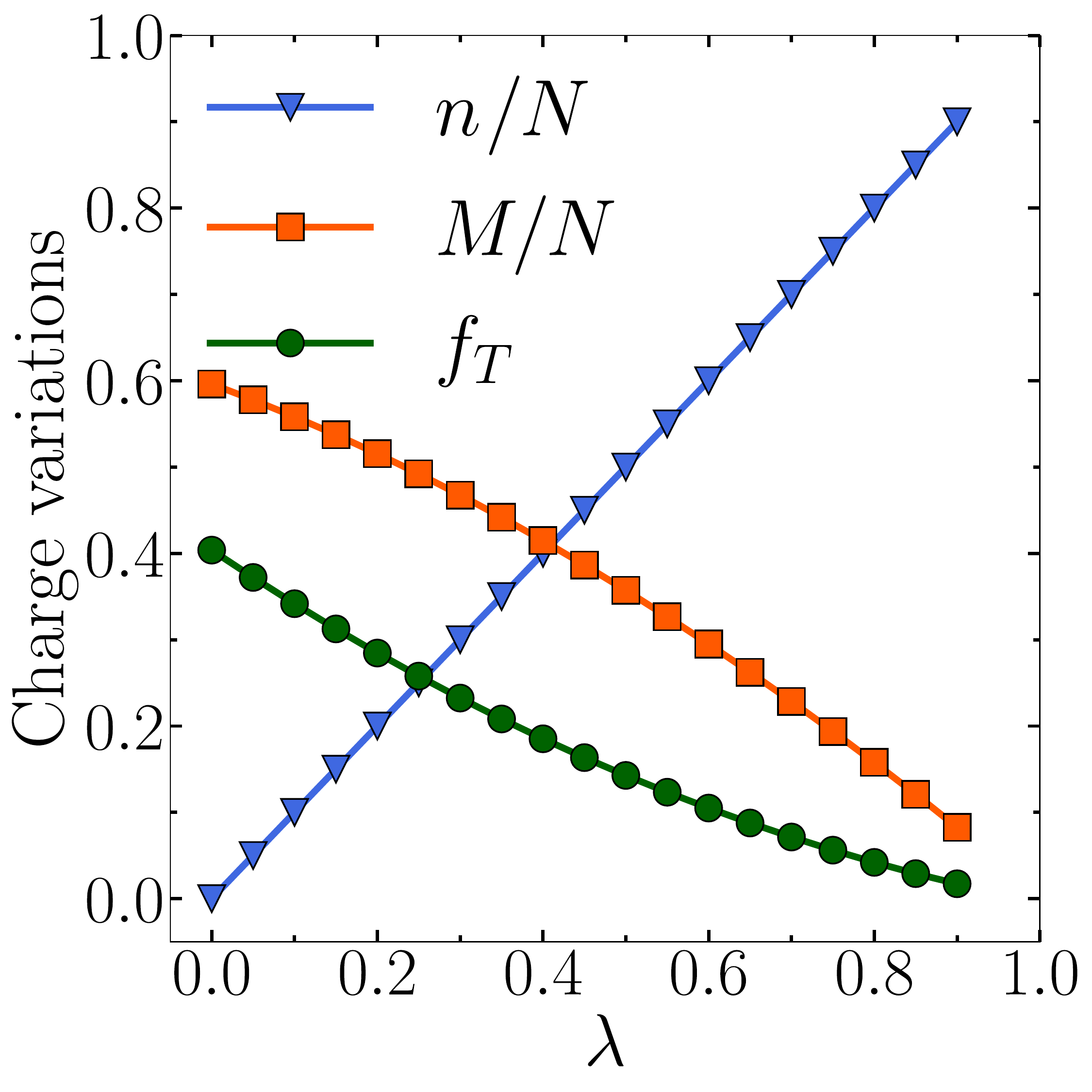}}%
	\qquad%
	 	\subfloat[]{\includegraphics[height=3.75cm,width=4.25cm]{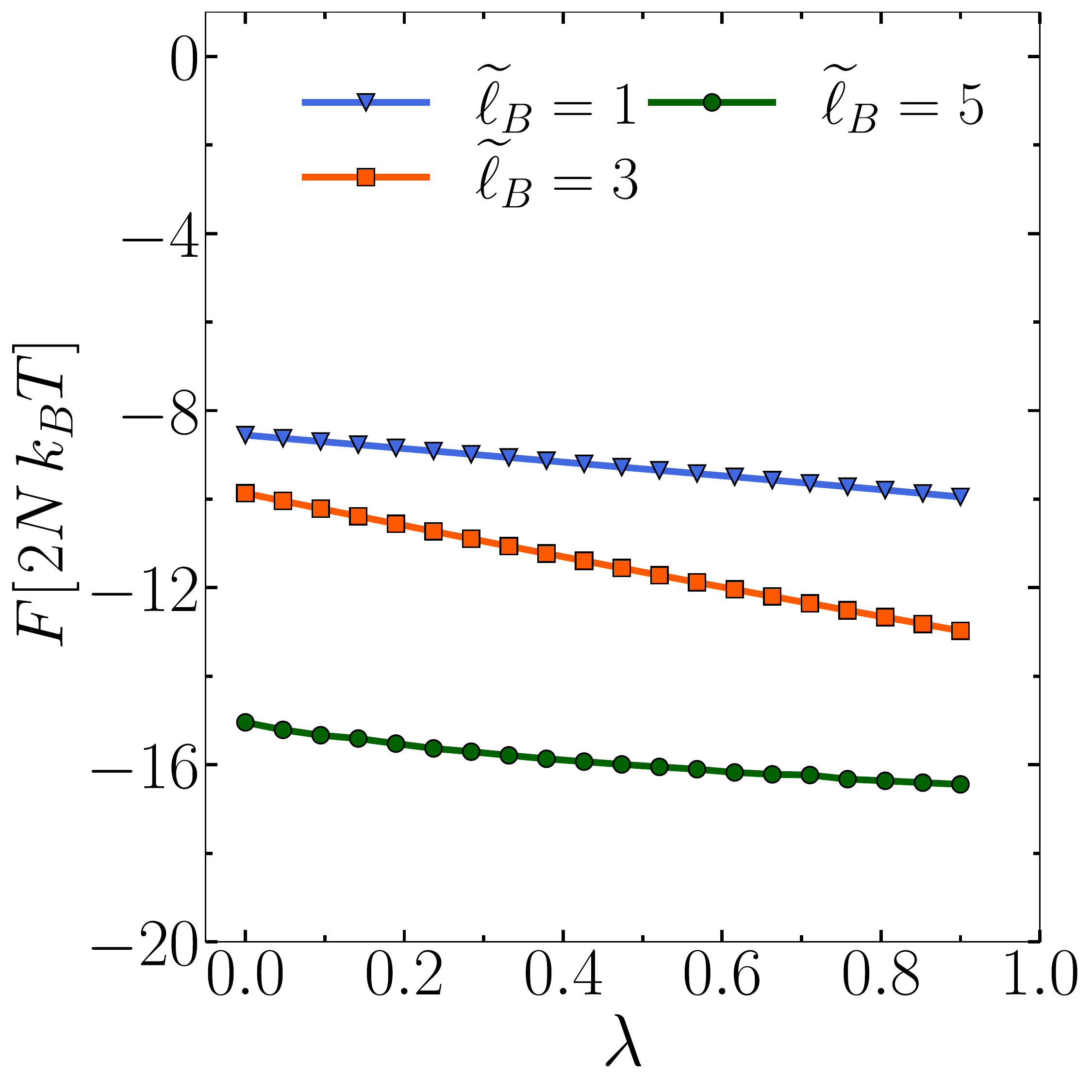}}
	 	\subfloat[]{\includegraphics[height=3.75cm,width=4.25cm]{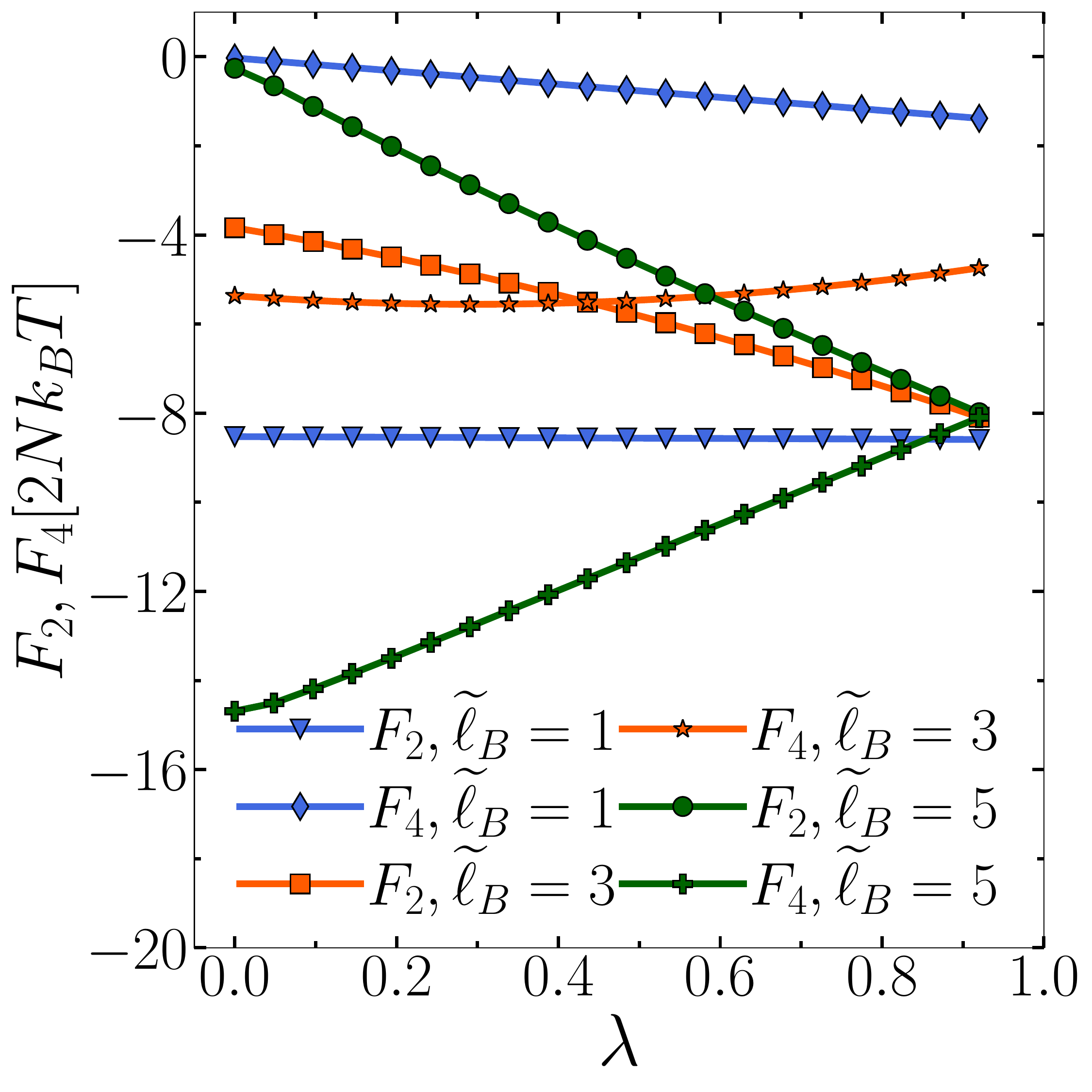}}
	\caption{Complexation of symmetric, fully ionizable polyelectrolytes (zero salt): (a) The size 
($\tilde{\ell}_1 \equiv \tilde{\ell}_{11}=\tilde{\ell}_{22}$) of the dangling (uncomplexed) parts and charge (total degree of ionization, $f_T \equiv f_{T1}=f_{T2}$) (inset)  of either chain (polycation/polyanion) are plotted as functions of degree of overlap $\lambda=n/N$ ($n,N$ are number of complexed and total monomers, respectively, at an intermediate state). 
(b) Total degree of ionization ($f_{T}$), degree of counterion condensation for the dangling parts ($M/N$), and degree of monomer complexation ($n/N$, fraction of monomers complexed) for either polyion are plotted against the degree of overlap $\lambda$. (c) The total free energy 
($F$) and (d) bound-pair enthalpy ($F_4$) and free ion entropy ($F_2$), all in units of $(N_1+N_2)k_B T=2Nk_B T$, are plotted against $\lambda$. For all four plots the parameters are: $N_1=N_2=N=1000, N_{c1}=N_{c2}=N=1000, \delta_1=\delta_2=\delta_{12} \equiv \delta=3.0, \tilde{\ell}_B=3.0$ [a) and b)], $w_{ij}=0.0, \tilde{c}_s=0.0$ and $\tilde{\rho}_i=0.0005$. Size and charge decrease with overlap. Complexation is enthalpy dominated at low, entropy dominated with nominal enthalpy change at moderate, and entropy dominated with substantial enthalpy loss at high Coulomb strengths. Free energy decreases linearly with overlap at modest Coulomb strengths.}%
	\label{fig:size}%
\end{figure}

 Both the size of the dangling parts ($\tilde{\ell}_1$) and charge (the total degree of ionization, $f_T$) of the symmetric PE chains monotonically fall [Fig. \ref{fig:size} (a)] with the overlap parameter ($\lambda \equiv n/N_{c1}=n/N$ for the symmetric case) as the complexation process proceeds in zero salt, releasing the condensed counterions, as well as giving way to formation of monomer-monomer ion-pairs in the complex 
[Fig. \ref{fig:size}(a) and (b)]. The complex is formed of fully and equally charged polyions, by displacing counterions with polymer segments [Fig. \ref{fig:size}(b)]. The degree of counterion condensation decreases, whereas the oppositely charged monomers make more pairs, resulting in a gradual decrease in the degree of ionization of the dangling chains ($f$), as well as the total degree of ionization ($f_{T}$), as overlap progresses. The final complex is neutral, and its stability is driven, for most moderate conditions, by the change in the enthalpy of bound ion-pairs and entropic free energy of free ions. To be specific, enthalpy gain dominates complexation equilibrium at low Coulomb strengths ($\tilde{\ell}_B=1$ for $\delta=3.0$), entropy gain dominates whereas enthalpy change is nominal at intermediate and moderate strengths ($\tilde{\ell}_B=3$), and finally enthalpy loss is substantial but entropy gain due to released counterions overcomes it to drive the process at high Coulomb strengths ($\tilde{\ell}_B = 5$) [Fig. \ref{fig:size}(c) and (d)]. For almost all Coulomb strengths which favour complexation, the dimensionless free energy per monomer, that is the free energy ($F$) in the units of $(N_1+N_2)k_B T=2Nk_B T$, versus overlap ($\lambda$) curve is linear with a negative slope [Fig. \ref{fig:size}(c)], consistent with the simulations\cite{peng2015,dzubiella2016}. The bound-pair enthalpy ($F_4$) and entropic contribution due to released counterions to the free energy ($F_2$) (in the same units of $F$) are plotted as a function of overlap for three different values of $\tilde{\ell}_B$ in [Fig. \ref{fig:size}(d)]. The free energy gain is maximum for moderate Coulomb strengths ($\tilde{\ell}_B=3$ for $\delta=3.0$), obtained previously\cite{mitra2023}.


	\subsubsection{Asymmetric polyelectrolyte chains: effect of unequal lengths}
	
Because of the asymmetry, the charge and size of the chains may not, in general, be taken equal for the two polyions in what follows in the paper. Therefore, the minimization of the free energy (Eq. \ref{Ftotal}), in general, needs to be carried out with respect to four variables, $f_{T1} \neq f_{T2}$ and $\tilde{\ell}_{11} \neq \tilde{\ell}_{12}$. 
	
\begin{figure}[!htbp]%
	 	\centering%
	 	\subfloat[]{\includegraphics[height=3.50cm,width=4.25cm]{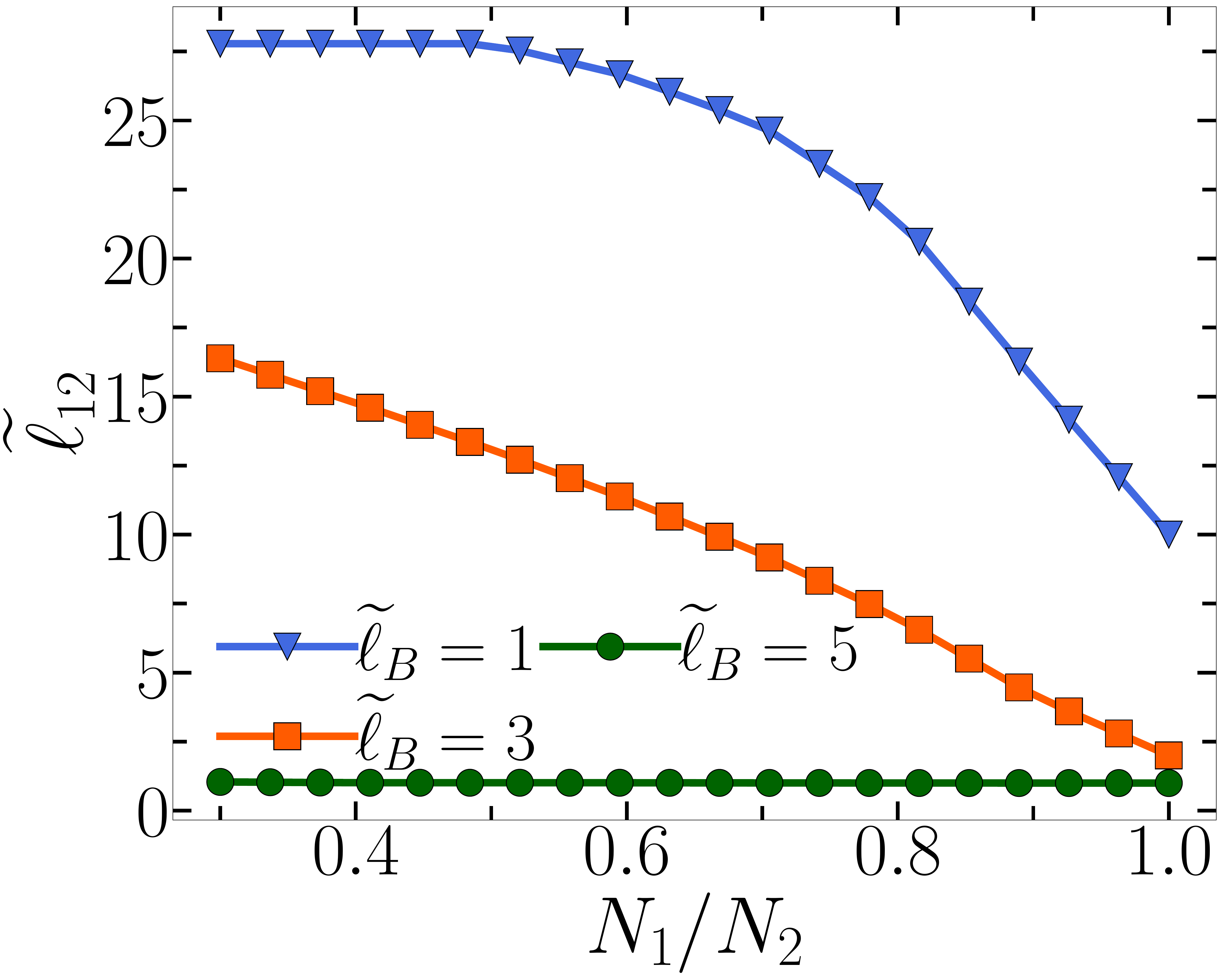}}
	 	\subfloat[]{\includegraphics[height=3.50cm,width=4.25cm]{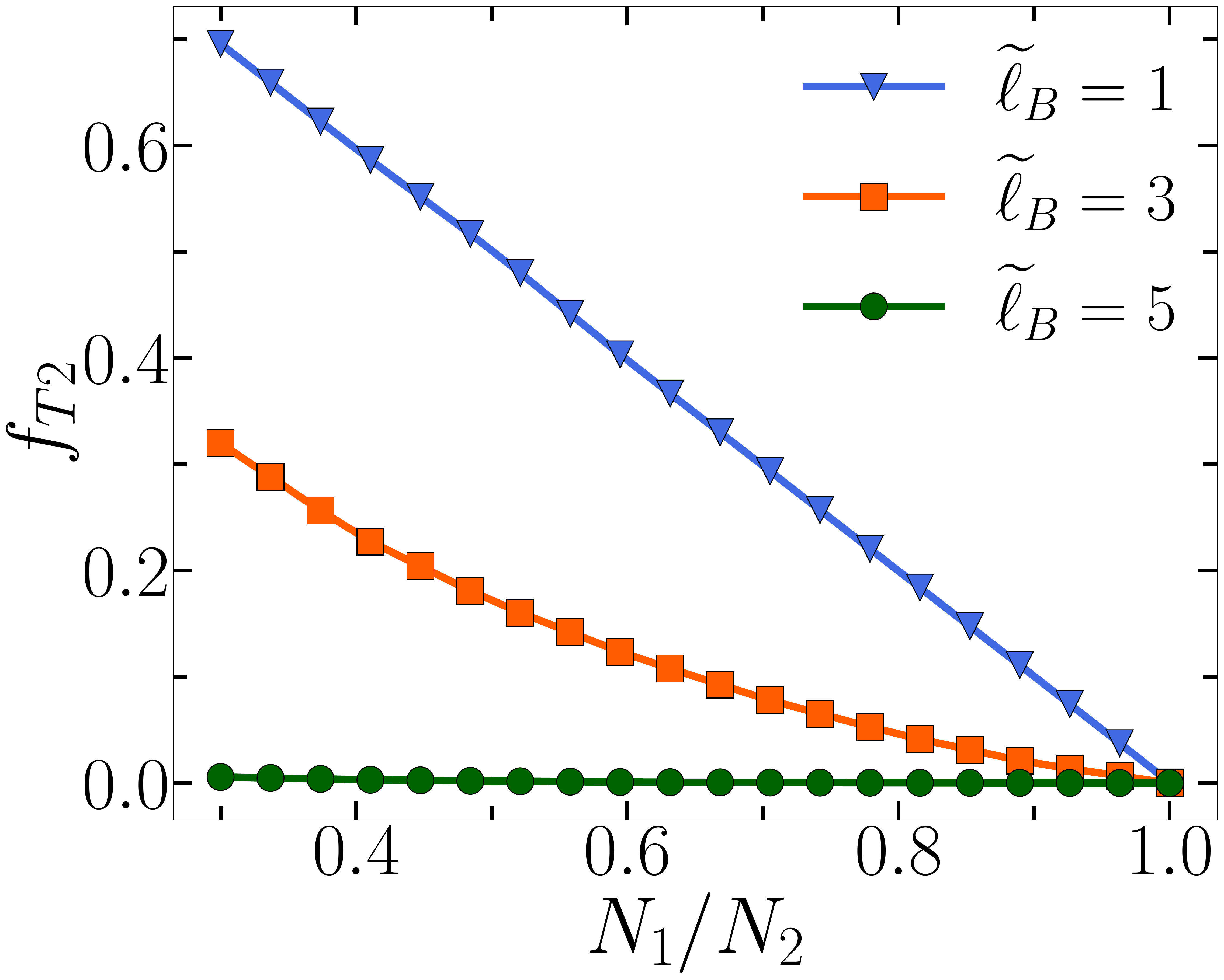}}
	 	\qquad%
	 	\subfloat[]{\includegraphics[height=4.00cm,width=5.25cm]{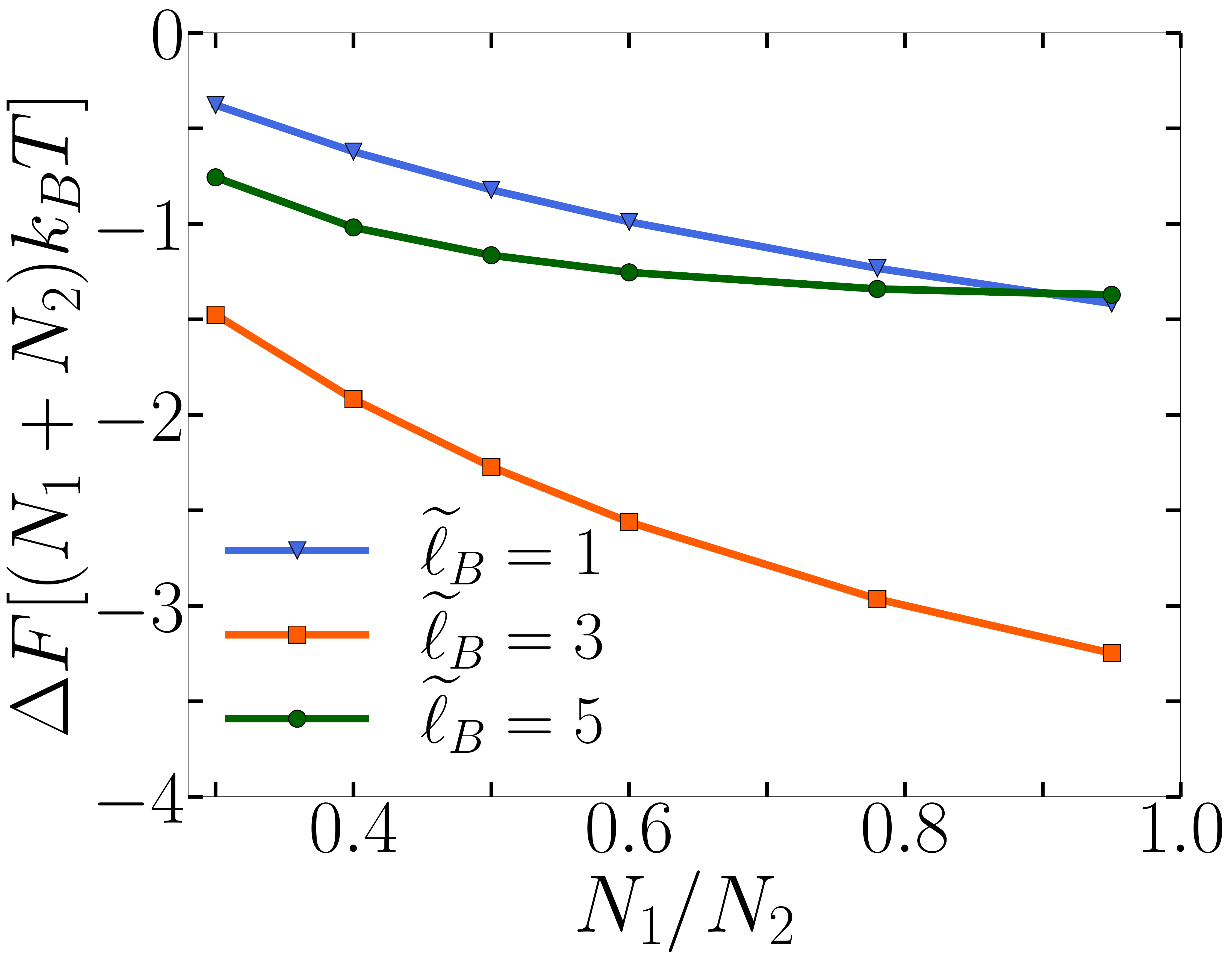}}
	 	\caption{Length asymmetry of complexing polyions: (a) The size ($\tilde{\ell}_{12}$) of the complex [or the longer chain (polycation, of length $N_2$)], (b) charge ($f_{T2}$) of the complex, and (c) free energy of complexation ($\Delta F$), in units of $(N_{1}+N_{2})k_BT$, is plotted against the asymmetry ratio ($N_1/N_2$) for different Coulomb strengths ($\widetilde{\ell}_{B}$). Both chains are fully ionizable ($f_{m1}=f_{m2}=1$). $N_1$ is varied from 0 to 1000 keeping $N_2=1000$. The parameters are: $\tilde{\rho}_2=0.0005$, $\tilde{\rho}_1$ is varying, $\tilde{\ell}_B=1,3,5, \delta_1=\delta_2=\delta_{12} \equiv \delta=3.0$, $w_{ij}=0.0$, and $\tilde{c}_s=0.0$. Longer chains form more stable complexes, the stability being maximum for moderate Coulomb strengths.}
	 	\label{fig:size-asy}
	 \end{figure}

	To see the effect of unequal lengths of the complexing polyions (asymmetric PE chains with $N_1 < N_2$, i.e., with PA smaller than PC, without any loss of generality), they are chosen to be fully ionizable ($N_{ci}=N_i$), the ratio of ionizable monomers ($N_{c1}/N_{c2}$) being equal to the ratio of lengths. No matter the degree of asymmetry, only two oppositely charged chains (that means only one smaller chain) are considered in this work.  After complexation, only a part of the charge of the longer chain ($f_{T2}$) is compensated, hence the chain is still expanded, dictating the size of the complex formed. To parameterize the asymmetry, we keep the length of the larger chain 
fixed at $ N_{2}=1000 $, and vary the smaller chain length, thereby introducing an asymmetry factor 
$ N_{1}/N_{2}$. The ratio $ N_{1}/N_{2}=1$ corresponds to the symmetric case. The 
degree of expansion decreases with less asymmetry, as the lengths of the chains become 
progressively equal leading to greater charge compensation. In Fig. \ref{fig:size-asy}(a-b), both the size and the charge of the complex 
($\tilde{\ell}_{12}$ and $f_{T2}$, respectively) fall monotonically as the system becomes more symmetric. The gain in free energy per monomer, $
\Delta F/(N_1+N_2)$ (in units of $k_B T$) increases with better symmetry, and is maximum for intermediate Coulomb 
strengths ($\delta=3.0, \tilde{\ell}_B=3.0$) for all degrees of asymmetry [Fig. \ref{fig:size-asy}(c)], which is consistent with the  results of the symmetric chains\cite{mitra2023}. The signature of the single chain properties applies to the asymmetric complex as well, as the size and charge fall with 
$\widetilde{\ell}_{B}$, counterion condensation being the underlying reason. For high Coulomb 
strengths ($\tilde{\ell}_B \sim 5.0$ for $\delta=3.0$), for all values of the asymmetry, the smaller chain 
and the counterions of the larger chain remain condensed on the larger chain. Overall, the 
bare charge of the larger chain ($f_{T2}$) remains very low. Accordingly, the size remains small and 
closer to the Gaussian conformation ($\tilde{\ell}_{12}=1.0$) or smaller. For lower Coulomb strengths, the counterions get released from the larger polyion, but the smaller chain remains complexed. For high asymmetry (low values of $N_1/N_2$), there are very few charged monomers from the smaller chain ($N_1$) to compensate the larger chain electrostatically. Therefore, the charge remains high, and so does the size for moderate to low Coulomb strengths as long as it is not too low (not lower than around $\tilde{\ell}_B=1.0$ for $\delta=3.0$). As 
the free energy is the lowest (or the gain is the highest) when $N_{1}=N_{2}$, it re-establishes 
the fact that the complex becomes more stable in the case of the symmetric chain lengths and at moderate values of $\widetilde{\ell}_{B}$ [for $\widetilde{\ell}_{B}=3.0, \delta=3.0 $, Fig. \ref{fig:size-asy}(c)]. At high $\tilde{\ell}_B$, the enthalpic loss due to the reduction in the number of bound ion-pairs lowers the free energy gain significantly\cite{mitra2023}. Studying multiple shorter chains would be an interesting proposition\cite{kabanov1985}, which is beyond the scope of this work.         
  
  A comment on the role of charged soluble complexes (the complexed polyion pairs discussed in this work) in bulk PE coacervation is in order. Coacervation is known to be a two-step process, namely the formation of soluble complexes with two chains followed by an assembly of such complexes which phase separate\cite{priftis2012,vitorazi2014,chang2017,adhikari2018,yethiraj2021,chen2022}. If the dangling part of the longer chain for these soluble complexes of unequal chain lengths remains charged (progressively more for lower values of $N_1/N_2$ in the above analysis), there is repulsion between them that hinders coacervation and stabilizes a dilute assembly of such complexes. In other words, the soluble complexes become hydrophilic\cite{kabanov1985} that may lead to steep decline in coacervate densities with lower charge\cite{neitzel2021}. However, such dangling chains individually attract counterions to reduce its average charge. Presence of salt will induce more counterion adsorpton, reduce the charge, and facililate coacervation in this complex interplay of charge neutralization and repulsion\cite{zhang2005,perry2014,friedowitz2021,bobbili2022,chen2022}. Increase of chain length and symmetry will increase the number of released counterions (seen in the two-chain simulations\cite{juarez2015}) and the free energy gain, which is evident from Fig. \ref{fig:size-asy}(c).
This in turn favors coacervation as seen in the literature\cite{adhikari2018,gucht2012,priftis2012,kayitmazer2015}. The charge content of the complexing chains, and its asymmetry either in the form of length or charge density, hence, have a major role to play in the structural and thermodynamic aspects of both soluble complex formation and coacervate phase separation.  
	 

	\subsection{Partially ionizable polyelectrolyte chains and complexation: effect of charge density and asymmetry}


In this subsection too, due to the asymmetry in either size or charge density of the chains PA and PC, the minimization of the free energy (Eq. \ref{Ftotal}) will be continued with respect to four variables, $f_{T1} \neq f_{T2}$ and $\tilde{\ell}_{11} \neq \tilde{\ell}_{12}$. As mentioned before, $\delta_1=\delta_2=\delta_{12} \equiv \delta$ and $w_{11}=w_{22}=w_{12}$ or $w_{ij}=0.0$ are assumed for the calculations. For the $i$-th chain of length $N_i$, the monomer density is given by $\tilde{\rho}_i =N_i/(\Omega/\ell^3)$. For $N_i=1000$, it turns out to be $\sim 0.0005$.

\subsubsection{Effective charge and size of the chains and the complex}

\begin{figure*}[!htbp]%
	\centering%
	\begin{tabular}{cc}
		\subfloat[]{\includegraphics[height=4.00cm,width=5.25cm]{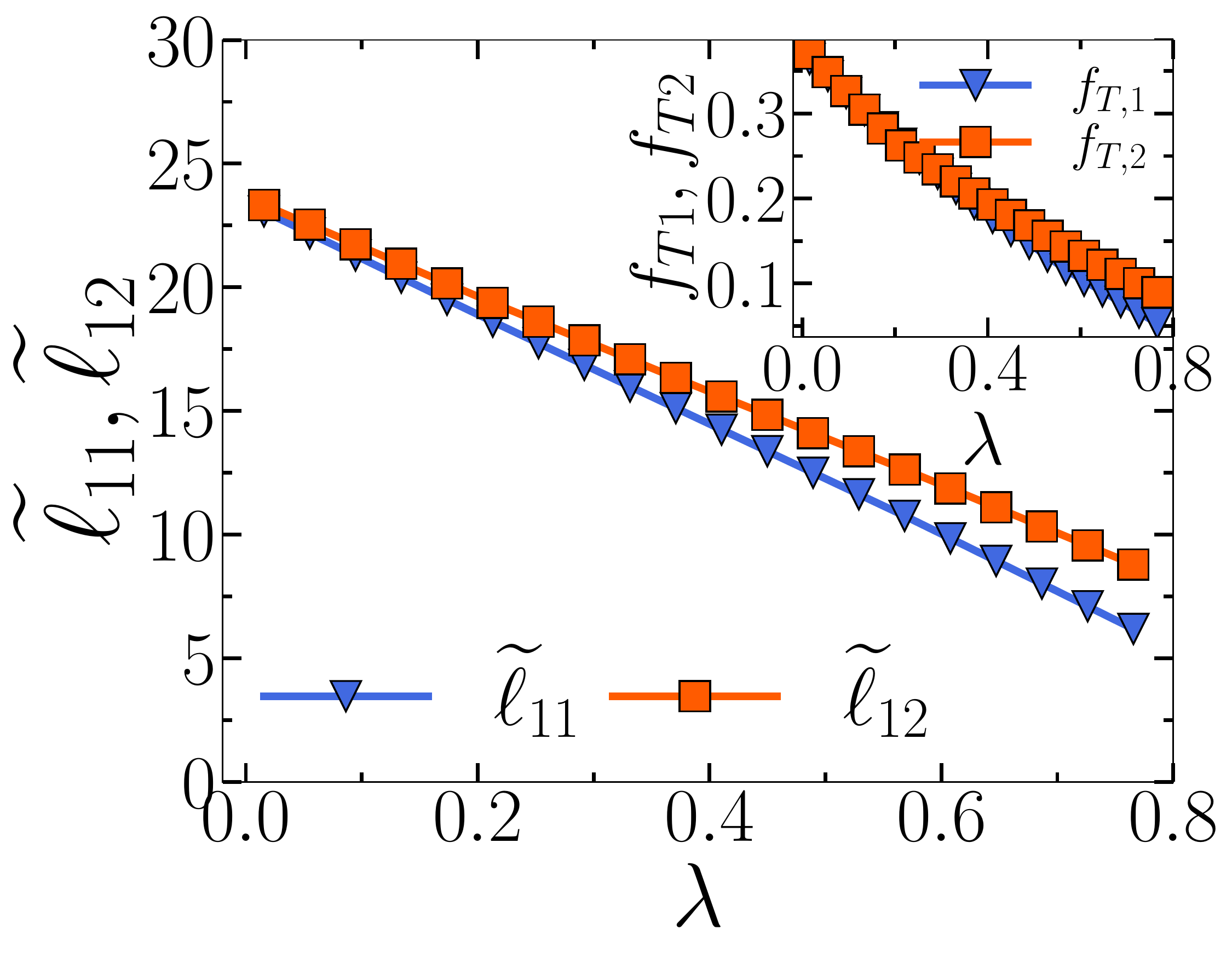}} & \subfloat[]{\includegraphics[height=4.00cm,width=5.25cm]{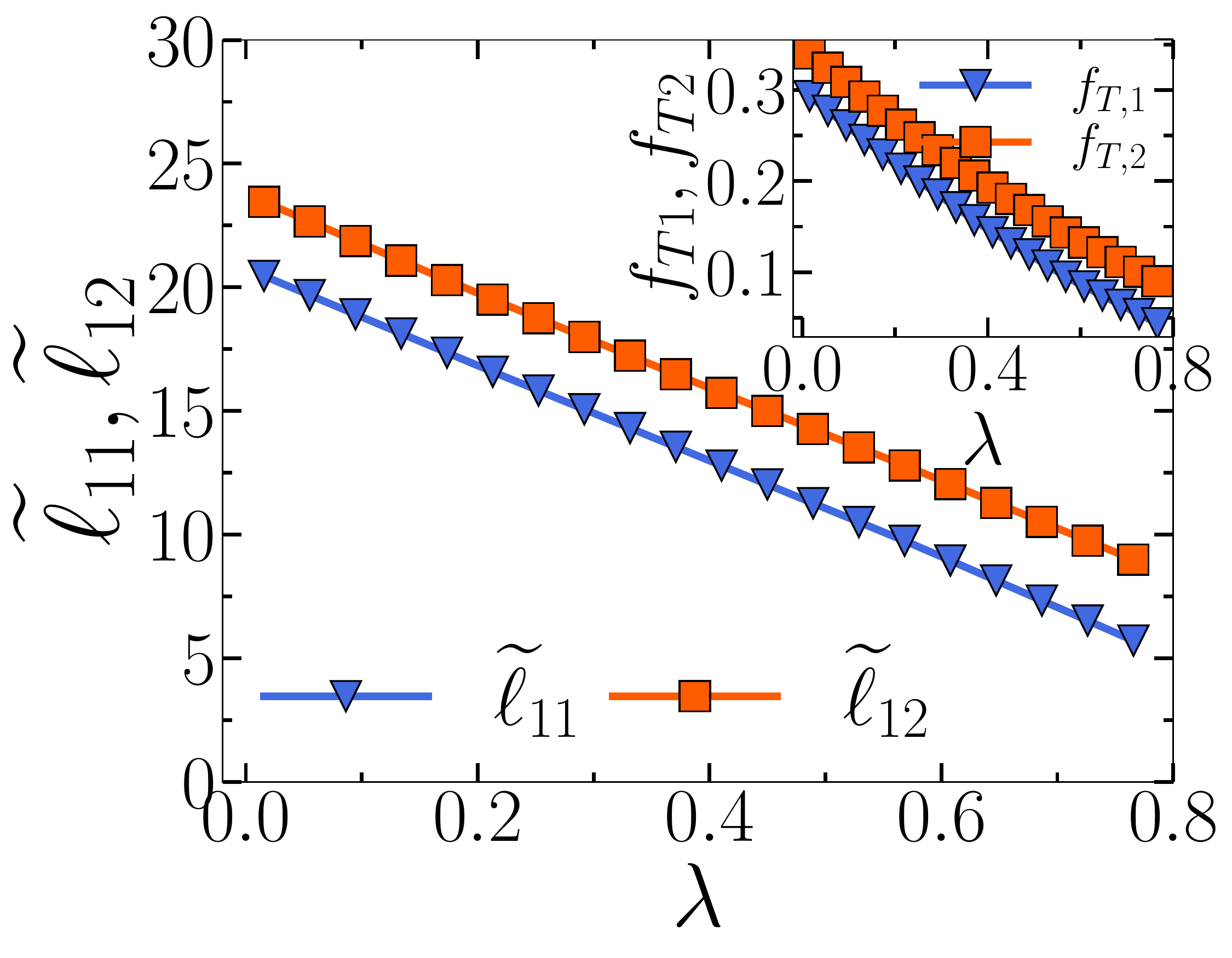}} \\
		\subfloat[]{\includegraphics[height=4.00cm,width=5.25cm]{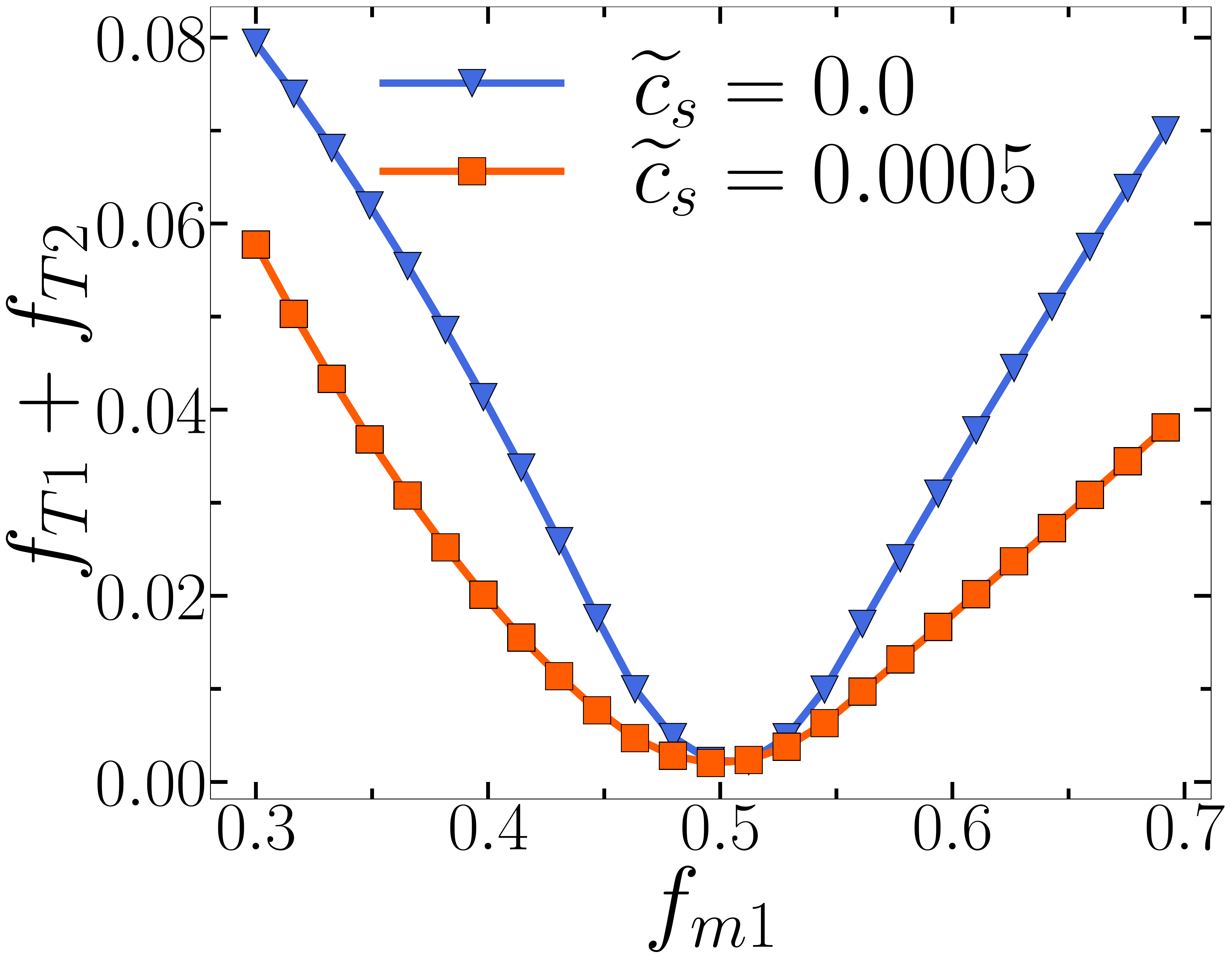}} & \subfloat[]{\includegraphics[height=4.00cm,width=5.25cm]{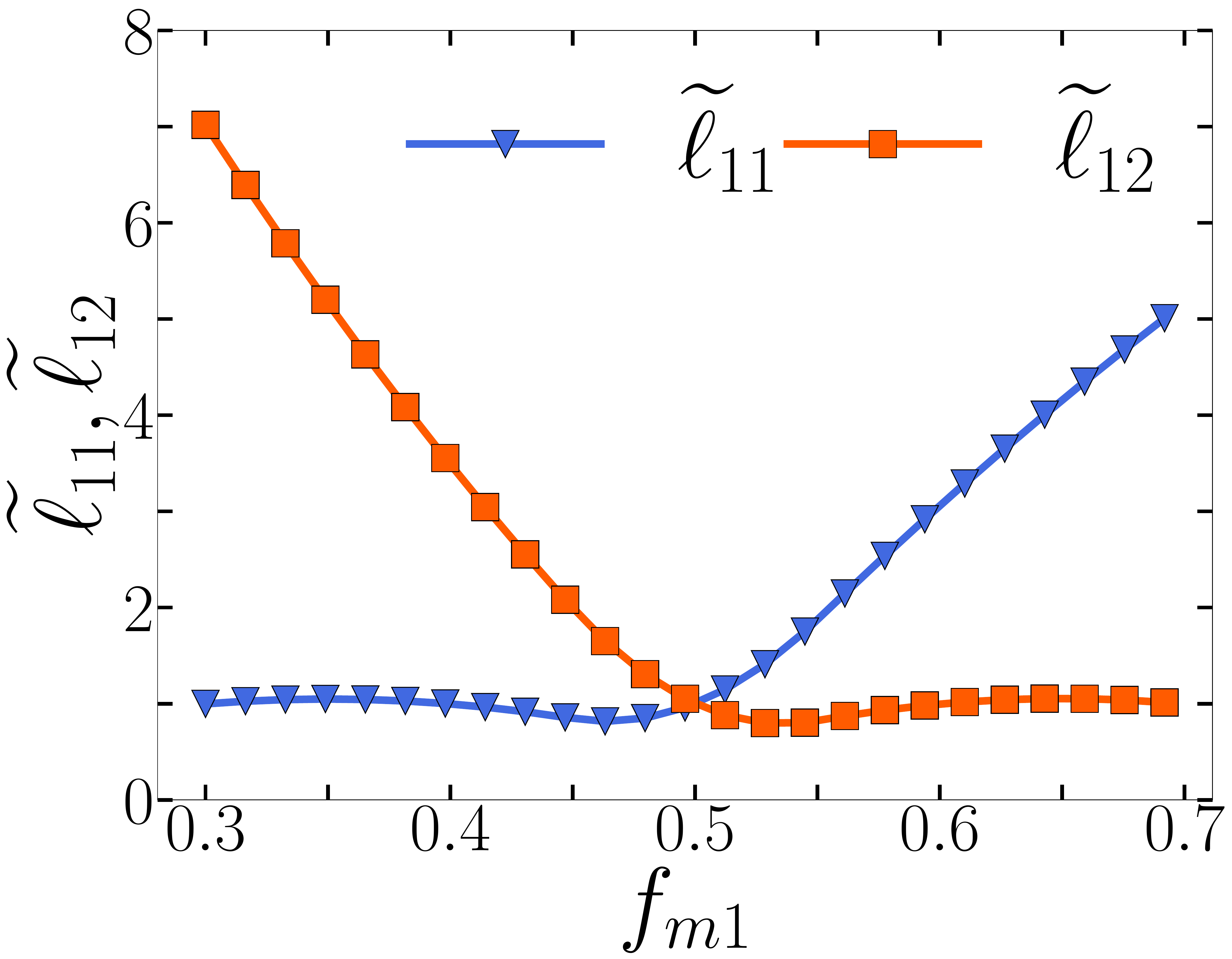}} \\
	\end{tabular}
	\captionsetup{justification=justified,margin=1cm}
	\caption{Charge asymmetry of complexing polyions: Sizes ($\tilde{\ell}_{11}$, $\tilde{\ell}_{12}$) and charges ($f_{T1}$, $f_{T2}$) (inset) as functions of overlap ($\lambda=n/N_{c1}$) of (a) polyanion with $N_1=800, N_{c1}=640$ and polycation with $N_2=1000, N_{c2}=800$ (so that $f_{m1}=f_{m2}=0.80$, fixed) and (b) polyanion with $N_1=1000,N_{c1}=640$, and polycation with $N_2=1000,N_{c2}=800$ (so that $N_1=N_2$ but $f_{m1}=0.64$ and $f_{m2}=0.80$) are plotted. Charge content of chains are kept the same. The parameters are: $\tilde{\ell}_B=3.0, \delta_1=\delta_2=\delta_{12} \equiv \delta=3.0, \tilde{c}_s=0.0$. (c) Net charge of the complex ($f_{T1}+f_{T2}$) and (d) size of the two PE chains ($\tilde{\ell}_{11}$ and $\tilde{\ell}_{12}$) are plotted against ionizability of the polyanion ($f_{m1} \equiv N_{c1}/N_1$) with $N_1=N_2=1000,N_{c2}=500$ ($f_{m2}\equiv N_{c2}/N_2=0.5 $), while $N_{c1}$ is varied from 300 to 700. The parameters are: $\tilde{\ell}_B=3.0,\delta_1=\delta_2=\delta_{12} \equiv \delta=3.0$, and two values of salt, $\tilde{c}_s=0.0$ and $\tilde{c}_s=0.0005$ are chosen. Other parameters are $w_{ij}=0.0$, $\tilde{\rho}_i=N_i/(\Omega/\ell^3)$ (e.g., $\tilde{\rho}_2=0.0005$ for $N_2=1000$). The complex becomes isoelectric (net charge zero) when both the chains have the same ionizability. With increasing charge difference the net charge of the complex and size also increase with an almost (but not completely) symmetric spread on the both sides of the isoelectric ionizability $ f_{m1}=0.5 $. }%
	\label{fig:size_par}%
\end{figure*}

Two symmetric, equally ionizable, and oppositely charged PE chains form a neutral complex at the mean field level\cite{zhaoyang2006,mitra2023,dzubiella2016,chen2022}. Such a complex may, however, end up with a non-zero net charge if there is charge asymmetry (stoichiometric difference), either due to difference in lengths or in charge contents of the chains. 
To study this issue, we looked into two different set ups, relevant for Figs. \ref{fig:size_par} (a,b). First, different chain lengths ($N_1 \neq N_2$) are taken but their ionizability is chosen to be the same ($f_{m1}=f_{m2}$). Second, the polyions in the complexing pair have the same length ($N_1 = N_2$) but different ionizabilities ($f_{m1} \neq f_{m2}$). The parameters are chosen in such a way that the charge content (total number of ionizable monomers) of the individual chains is the same in the two set ups ($N_{c1} \equiv f_{m1}N_1$ is the same for set ups 1 and 2, and so is $f_{m2}N_2 \equiv N_{c2}$). 

The sizes ($\tilde{\ell}_{1i}$) and total degrees of ionization ($f_{Ti}$) [inset] of polyanion ($ N_{1}=800,N_{c1}=640,f_{m1} =0.8$) and polycation ($ N_{2}=1000,N_{c2}=800,f_{m1} =0.8$) are plotted in Fig. \ref{fig:size_par}(a) for set up 1 and of polyanion ($ N_{1}=1000,N_{c1}=640,f_{m1} =0.64$) and polycation ($ N_{2}=1000,N_{c2}=800,f_{m1} =0.8$) in Fig. \ref{fig:size_par}(b) for set up 2, as functions of the overlap of the chains ($\lambda=n/N_{c1}$, where $n$ is the number of monomer-monomer ion-pairs in an intermediate step of the complex). As the number of ionizable monomers (the total chemical charge) in PA is chosen either smaller or equal to that of PC ($N_{c1} \le N_{c2}$),
 the complexation is complete when $n=N_{c1}$, or $\lambda=1$. The chain with higher charge density on its backbone remains relatively more swollen in size throughout the complexation process, and the absolute values of the sizes and charges are quantitatively similar for the two set ups. This implies that it is not the ionizabilities ($f_{mi}$) but the charge contents (number of ionizable monomers, $N_{ci}$, or the charge stoichiometry) which are the key factors to dictate complexation of a pair of asymmetric, charged polyions. The degree of ionization does not vanish for the PE chain with higher chemical charge (PC), but the chain with lower chemical charge releases all the counterions in solution to become fully complexed. It is interesting to note from Fig. \ref{fig:size_par}(a-b) that the sizes of the chains are correlated in a way, so that the reduction in size of one chain automatically reduces the size of the other\cite{zhaoyang2006,peng2015,mitra2023}. This is expected since the complex being formed makes both the chains progressively lose out on their ionizable monomers, reducing the intra-polyion electrostatic repulsion that leads to reduction in size. Concurrently, the degree of ionization (or, the effective charge after charge compensation of the monomers) of the chains ($f_{Ti}$) also decreases progressively.

To investigate further the issue of asymmetry in charge density, we choose two polyions of the same length ($N_1=N_2=1000$) and vary the asymmetry in the chemical charge density in the form of ionizability ($f_{mi}$). For PC the ionizability is fixed at $f_{m2}=0.5$ $(N_{c2}=500)$, and for PA the ionizability ($f_{m1}$) is varied from 0 to 1 ($N_{c1}$ from 0 to 1000). In this case, we calculate the charge and size of fully complexed polyions ($\lambda=1$). Note that $\lambda=n/N_{c1}$ when $N_{c1} < N_{c2}$, but $\lambda=n/N_{c2}$, when $N_{c1} > N_{c2}$. Plotted in Fig. \ref{fig:size_par}(c), the effective net charge of the complex (the surviving uncomplexed charge of the polyion that has higher ionizability, given by either $f_{T1}$ or $f_{T2}$) is seen to be zero, expectedly, when the ionizability of PA is 0.5, equal to that of PC. Depending on which side of 0.5 the ionizability of PA is, the size of the complex is calculated in the mean field from the resultant ionizability of it (equal to $f_{m1}-f_{m2}$, negative or positive when $f_{m1}$ is less than or greater than 0.5, respectively), assuming that the resultant ionizable monomers outside the complexed part are distributed over all monomers of the chain ($N_i=1000$), including its complexed part. As the polyions are oppositely charged, in Fig. \ref{fig:size_par}(c), the modulus of such resultant charge ($f_{T1}+f_{T2}$) is plotted. It decreases in the presence of added salt, as due to reduced chemical potential there is more counterion condensation on the bare charges (which fail to be a part of the complexed part) of the chain with higher ionizability. Neither the charge nor the size are exactly symmetric (but close to it) for values of $f_{m1}=0.5+$ and 0.5- (equally higher and lower from the compensation point $ N_{c1}=N_{c2} $), because indeed the two cases are not electrostatically identical. For example, a pair with $N_{c1}, {N_1} = 300,1000$ and $N_{c2}, {N_2} = 500,1000$ and another with $N_{c2}, {N_2} = 500,1000$ and $N_{c1}, {N_1} = 700,1000$ will not be identical, because the degree of counterion condensation, no matter how marginally, would still be different in the two cases. At the compensation point the complex is comprised only of polyion monomers, after releasing all counterions, but moving away ($N_{c1} \neq N_{c2}$), the complex contains counterions as well.

As for the size ($\widetilde{\ell}_{11}, \widetilde{\ell}_{12}$), the PE with the lower chemical charge content assumes closer to (actually, smaller than) Gaussian conformation while the one with the higher charge content remains in a swollen configuration, the degree of swelling increasing with higher net charge [Fig. \ref{fig:size_par}(d)]. As mentioned for complexes formed of polyions of asymmteric length, the charge asymmetric complexes mentioned above with its resultant charge will hinder coacervation, or make the coacervates fluidic\cite{chen2021}, eventually having the lowest size and mutual repulsion for 1:1 charge stoichiometry that is the most favorable to form coacervates
\cite{voorn1957,hayashi2003,hayashi2004,chollakup2010,vitorazi2014,adhikari2018,bobbili2022,chen2022,zhang2005,gucht2012,
priftis2012,perry2014,neitzel2021} with the highest rigidity modulus\cite{porcel2009}. The size and repulsive nature of the polyion with higher ionizability in the complex, along with elevated structural free volumes of the coacervates made of such complexes\cite{chen2021}, decreases with salt due to counterion condensation and screening, leading to favorable coacervation\cite{zhang2005,friedowitz2021,bobbili2022,chen2022} or decreasing porosity of an already formed coacervate\cite{porcel2009}. The variation in the chemical charge density of one polyion keeping the other fixed [Fig. \ref{fig:size_par}(c-d)] may be compared to the tuning of the charge ratio in weak polyelectrolytes, by changing the pH of the solution\cite{lalwani2021,knoerdel2021,salehi2015,digby2022} resulting in controlling the `saloplasticity'.


\subsubsection{The driving forces of complexation: enthalpy or entropy? A diagram of states}

From general observation, three distinct regimes depending upon the driving entity, the free ion entropy or ion-pair Coulomb energy, may be associated with the pair-wise complex formation of two oppositely charged polyions in an implicit solvent. For low Coulomb strengths ($\delta \tilde{\ell}_B$), the complexation drive is dominated by the enthalpy gain of monomer-monomer ion-pairs. For moderate Coulomb strengths, the drive is dominated by entropy gain of free counterions, but with nominal change in enthalpy\cite{zhaoyang2006,laugel2006,fu2016,mitra2023,whitmer2018macro}. Finally, for high electrostatic strengths, the complexation is still driven by entropy gain but significantly opposed by enthalpy loss\cite{lundbck1996,matulis2000,bronich2001,zhaoyang2006,laugel2006,whitmer2018macro,
mitra2023}. 

\begin{figure*}[!htbp]%
		\centering%
		\subfloat[]{\includegraphics[height=4.30cm,width=5.25cm]{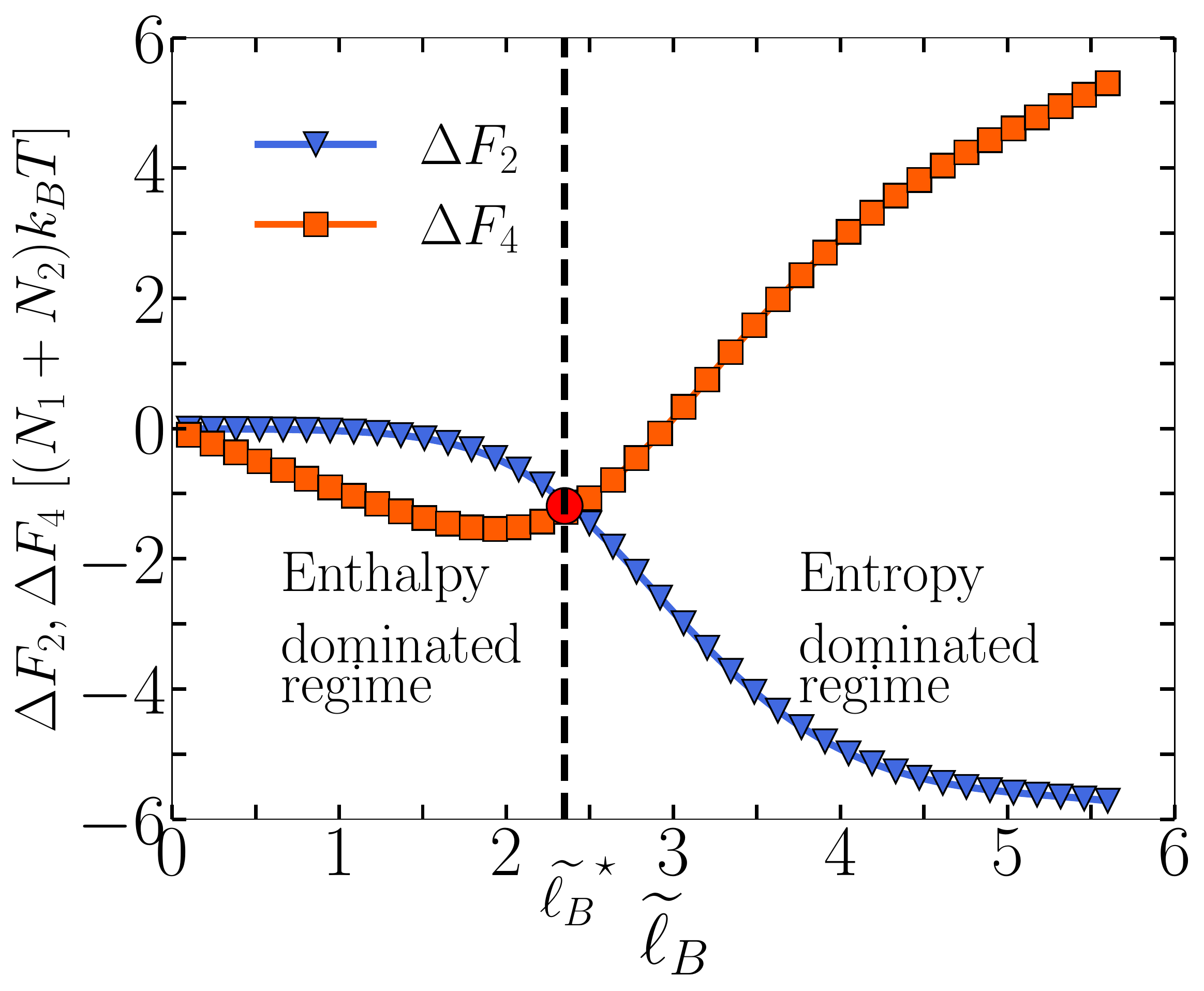}}\label{300_400}%
		\subfloat[]{\includegraphics[height=4.30cm,width=5.25cm]{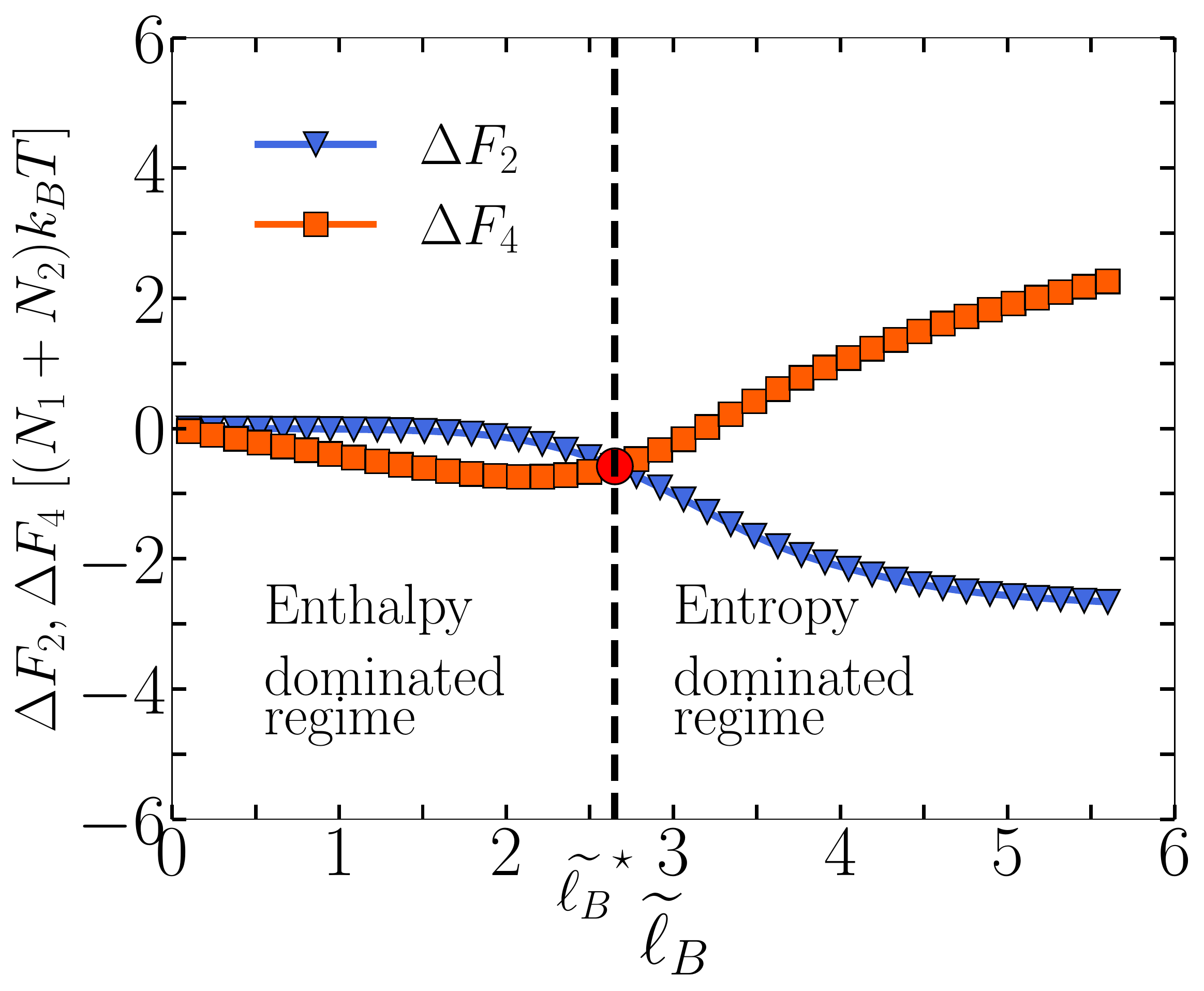}}\label{800_900}%
		\qquad%
		\subfloat[]{\includegraphics[height=4.00cm,width=5.25cm]{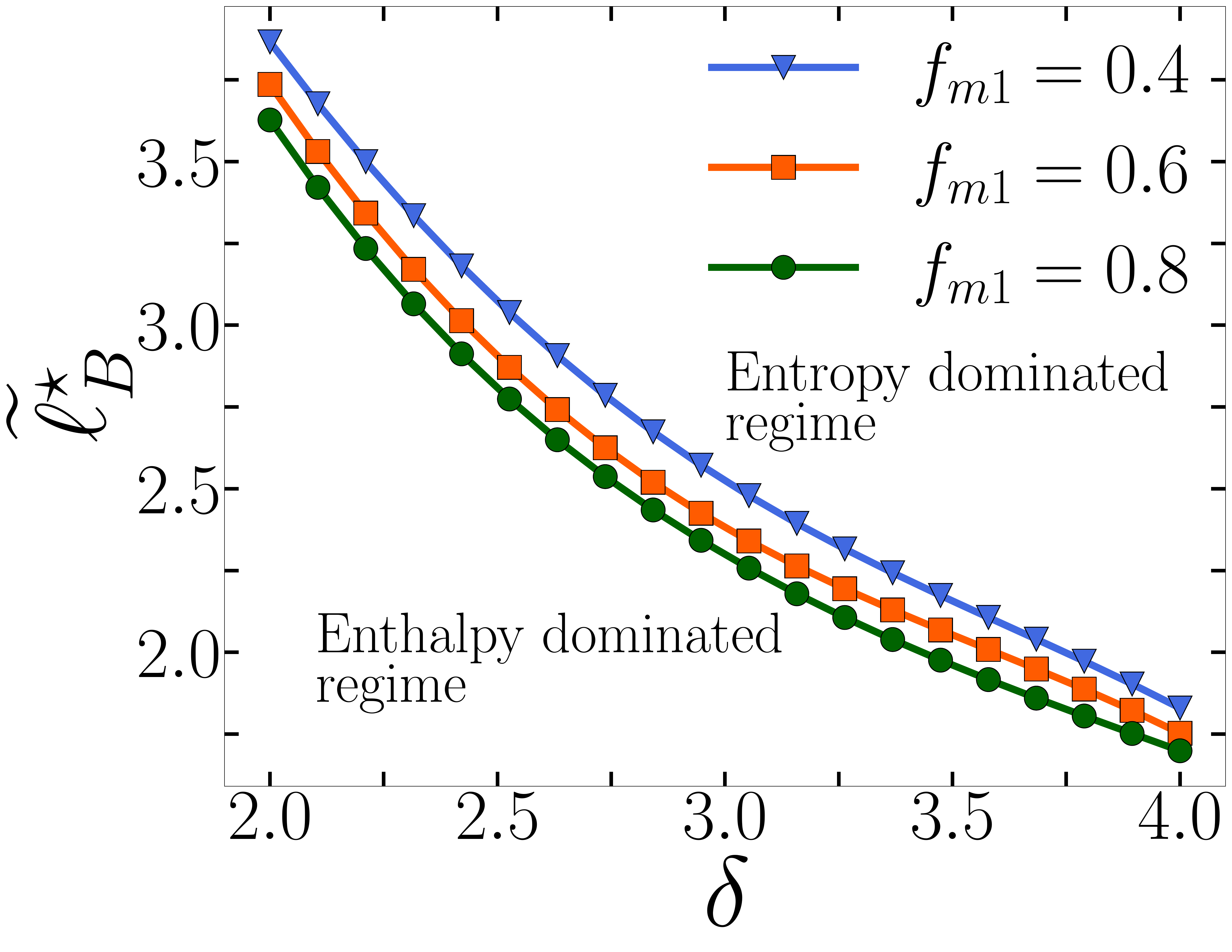}}
		\subfloat[]{\includegraphics[height=4.00cm,width=5.25cm]{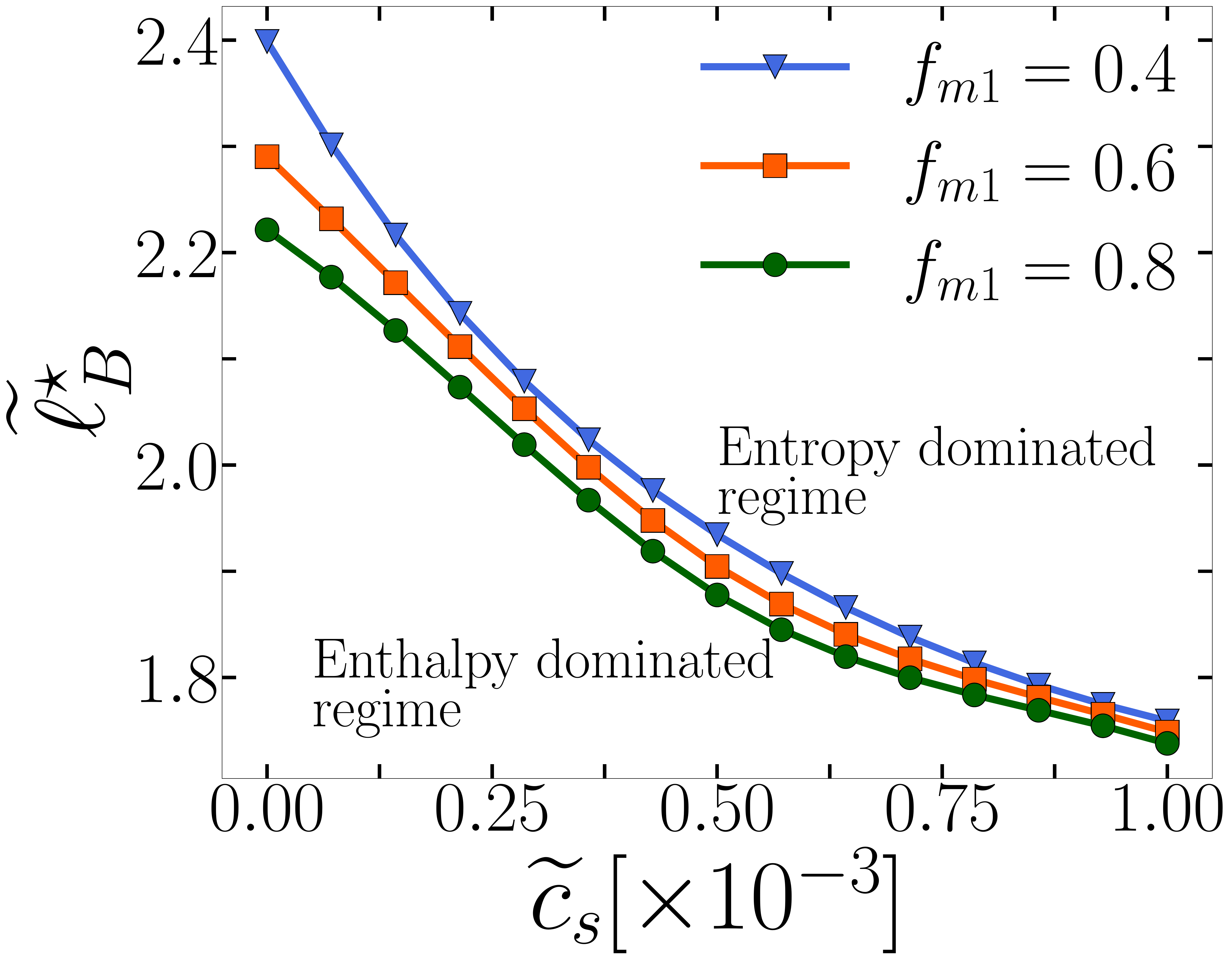}}
		\caption{Relative drive of enthalpy and entropy (major parts) of complexation: Variation in the free energy parts of complexation, $\Delta F_2$ (free ion entropy) and $\Delta F_4$ (ion-pair energy), in units of
		$(N_1+N_2)k_BT$, and as functions of Coulomb strength $\tilde{\ell}_B$ are plotted for (a) $N_1=N_2=1000, N_{c1}=N_{c2}=700$ ($f_{m1}=f_{m2}=0.7$) and (b) $N_1=N_2=1000, N_{c1}=N_{c2}=300$ ($f_{m1}=f_{m2}=0.3$). The crossover point, $\tilde{\ell}_B^\star$ (at which $\Delta F_2= \Delta F_4$), is identified. Other parameters are: $\tilde{c}_s=0.0, \delta_1=\delta_2=\delta_{12} \equiv \delta =3.0$, $w_{ij}=0.0$, and $\tilde{\rho}_i=0.0005$. The crossover strength $\tilde{\ell}_B^\star$ divides the complexation process in two regimes, the enthalpy-dominated to its left and the entropy-dominated to its right. For higher Coulomb strengths, the complexation becomes hindered by the enthalpy loss, but still driven by entropy gain. Variation of the crossover Bjerrum length $\tilde{\ell}_B^\star$ as (c) function of $\delta$ (keeping $\tilde{c}_s=0.0$), and (d) function of $\tilde{c}_s$ (keeping $\delta=3.0$) are shown for three values of ionizability $f_{m1}=0.4,0.6,0.8$ (keeping $N_{c1}=N_{c2}$ with $N_1=N_2=1000$, or $f_{m1}=f_{m2}$). Other parameters are: $\delta_1=\delta_2=\delta_{12} \equiv \delta =3.0$, $w_{ij}=0.0$, and $\tilde{\rho}_i=0.0005$ for (c) and (d), respectively. $\tilde{\ell}_B^\star$ shifts marginally to a lower value for higher ionizability of the polyions. However, it decreases significantly with both higher Coulomb strength initiated by $\delta$ (notable is how the product $\delta \widetilde{\ell}_{B}^\star$ stays almost constant, between $ \sim 7 $ to $ \sim 8 $) or higher salt.}\label{enthalpy-entropy}
	\end{figure*}

In Fig. \ref{enthalpy-entropy}(a-b) we plot, in units of $(N_1+N_2) k_B T$, the free energy of complexation due to the entropy of free ions [$\Delta F_2=F_{2}(\mbox{final})-F_{2}(\mbox{initial})$] and enthalpy of bound ion-pairs [$\Delta F_4=F_{4}(\mbox{final})-F_{4}(\mbox{initial})$]. At low Coulomb strengths, the polyelectrolytes are highly charged, and all the counterions from both chains remain released even before the onset of complexation. This disallows entropy gain due to released counterions at complexation, and the complex formation is solely driven by the electrostatic enthalpy of the system. With an increase in Coulomb strength, counterions start to adsorb on the chains before the onset of complexation, and get released during complexation. Hence, the entropy gain becomes progressively more relevant with higher Coulomb strengths. Eventually we find an intermediate temperature, $T^{\star}$ (or an equivalent condition corresponding to the crossover Coulomb strength, $\widetilde{\ell}_{B}^{\star} \sim 1/T^{\star}$, for a fixed $\delta$), at which the entropic and enthalpic drives for complexation become equally competitive, and beyond which the enthalpy dominance gives way to entropy dominance [Fig. \ref{enthalpy-entropy}(a-b)]. As $\Delta F_2$ increases with the number of counterions released, which in turn depends on the counterions condensed on the isolated chains, there is a striking similarity between $\Delta F_2$ versus $\tilde{\ell}_B$ and $f$ (degree of ionization) versus $\tilde{\ell}_B$\cite{muthu2004} curves. The crossover value of the Coulomb strength will in principle depend on the size ($N_1, N_2$) and number of ionizable monomers ($N_{c1}, N_{c2}$) of the complexing polyions, the dielectric parameter ($\delta$), and salt ($\tilde{c}_s$). For both Figs. \ref{enthalpy-entropy}(a) and (b), separately, the complexing polyions are of the same size ($N_1=N_2=1000$), and so are their number of ionizable monomers ($N_{c1} = N_{c2}$), hence the ionizabilities ($f_{m1} = f_{m2}$).  However, the ionizability is chosen to be different for cases (a) and (b), to emphasize on the electrostatic nature of the drive. For (a) $N_{c1}=N_{c2}=700$ and for (b) $N_{c1}=N_{c2}=300$, which reflect in a lower value of $\Delta F_2$ and $\Delta F_4$ and their variation in the latter. 

In Figs. \ref{enthalpy-entropy}(c) and (d) the crossover $\tilde{\ell}_B^\star$ is plotted against 
$\delta$ for fixed $\tilde{c}_s (=0.0)$, and against $\tilde{c}_s$ for fixed $\delta (=3.0)$, 
respectively, for different ionizabilities ($f_{mi}$). We note that for fixed $\delta$ and salt ($
\tilde{c}_s=0$), $\tilde{\ell}_B^\star$ decreases marginally with the charge densities ($f_{mi}$) of 
the chains. However, it decreases sharply with $\delta$ [Fig. \ref{enthalpy-entropy}(c)] or salt 
concentration [Fig. \ref{enthalpy-entropy}(d)]. The total crossover Coulomb strength ($\delta 
\tilde{\ell}_B^\star$) remains almost constant (between 7 and 8) for all values of $\delta$. For the 
variation in $\tilde{c}_s$, ($ \tilde{c}_s \tilde{\ell}_B^\star$) shows a very similar trend for all 
values of $\tilde{c}_s$. However, $\delta \tilde{\ell}_B^\star$ or $ \tilde{c}_s \tilde{\ell}_B^\star$ decreases marginally with ionizability $f_{mi}$. Both $\delta$ and $ \tilde{c}_s$ sensitively induce counterion condensation before complexation (for isolated chains), which in turn determines whether the complexation is enthalpy or entropy driven. This is the reason the dominance in drive, whether enthalpic or entropic, does change nominally with the charge density (ionizability) of the chains. The magnitude of such drives, of course, depends sensitively on the charge densities of such polyion pairs, and increases with it, as seen in Figs. \ref{enthalpy-entropy}(a-b).  

A more detailed analysis is in order to explain the marginal change in $\delta \tilde{\ell}_B^\star$ with $f_{mi}$. It is known that the effective Coulomb strength $\delta \tilde{\ell}_B$ critically determines the charge contents in polyelectrolytes (see, for example, Eq. \ref{single-chain-charge-anal}, wherein $\delta \tilde{\ell}_B$ features in the arguments of the exponentials). Therefore, marginal change in the crossover point $\delta \tilde{\ell}_B^\star$ with $\delta$, for fixed $f_{mi}$, is not surprising. It implies that the degree of counterion condensation $\alpha_i$ before the onset of complexation depends overwhelmingly on the product $\delta \tilde{\ell}_B$ (Eq. \ref{single-chain-charge-anal}), and as the enthalpy gain (and the entropy gain, which grows at the expense of a decreasing enthalpy gain) depends on $\alpha_i$, the crossover point $\delta \tilde{\ell}_B^\star$ at which the gains are equal is almost a constant even if $\delta$ and $\tilde{\ell}_B^\star$ vary individually. What is surprising is that $\delta \tilde{\ell}_B^\star$ does 
not change much with $f_{mi}$, which implies that the degree of counterion condensation, if one
considers the ionizable monomers only, does remain fixed with $\delta \tilde{\ell}_B$. In other words,
the quantity $f_{ci}=(N_{ci}-M_i)/N_{ci}=1-M_i/N_{ci}$ (defined as the degree of ionization with
respect to ionizable monomers only) does change marginally with $f_{mi}$ if $\delta \tilde{\ell}_B$ is fixed, leading to a marginal change in the crossover Coulomb strength $\delta \tilde{\ell}_B^\star$
with $f_{mi}$. Indeed, $f_{ci}$ is found to be around 0.75-0.80 for $\delta \tilde{\ell}_B^\star \sim 7-8$ for all values of $f_{mi}$. Therefore, $f_{ci}$ seems to be a constant of the problem for a fixed salt.      

\subsubsection{Entropy, enthalpy, and free energy of complexation near crossover Coulomb strength: effect of charge density}

\begin{figure}[!htbp]%
		\centering%
		\subfloat[]{\includegraphics[height=5.25cm,width=7.00cm]{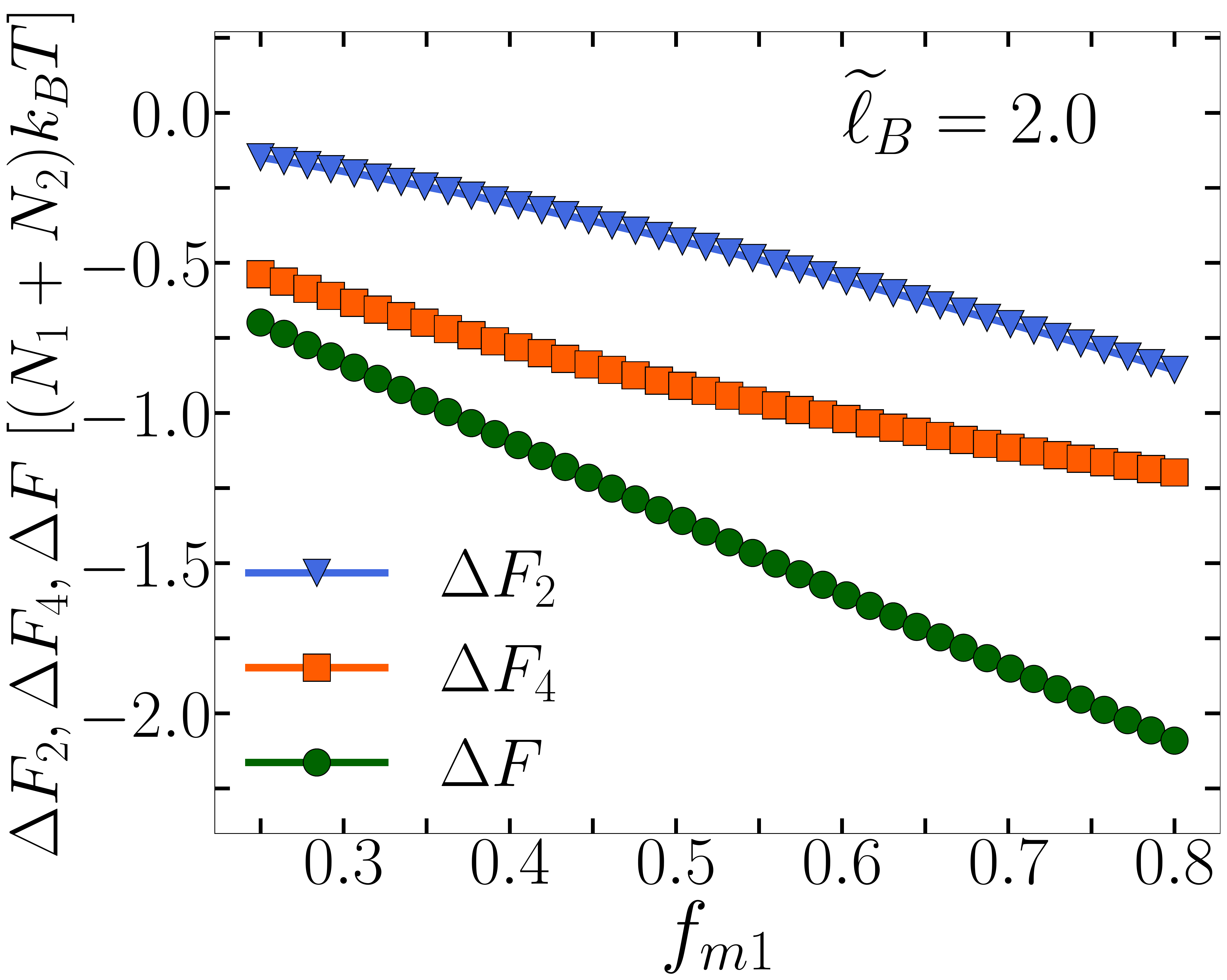}}%
		\qquad%
		\subfloat[]{\includegraphics[height=5.25cm,width=7.00cm]{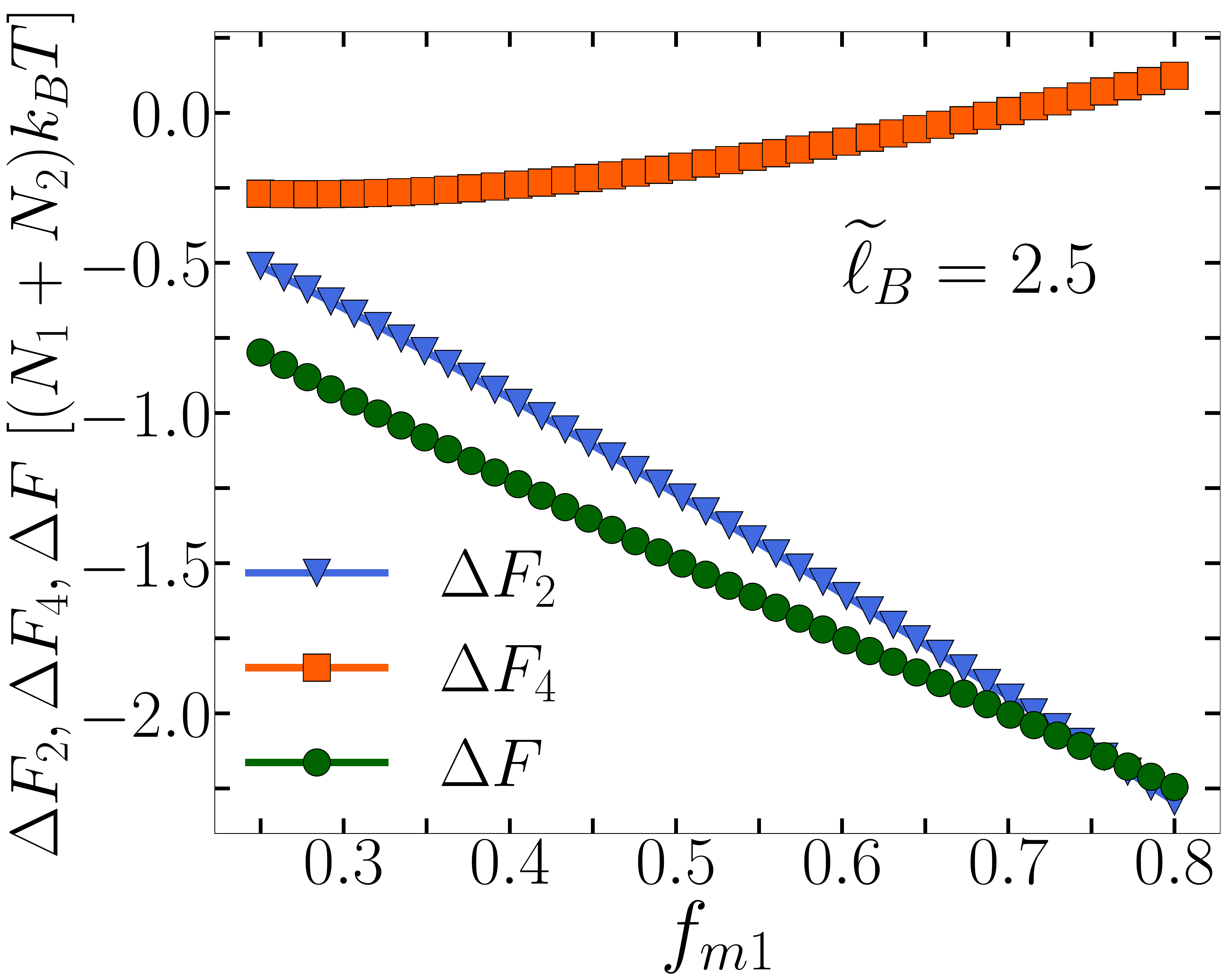}}%
		\qquad%
		\caption{Relative drive of complexation as function of ionizability: Variation in the free energy parts of complexation $\Delta F_2$ (free ion entropy) and $\Delta F_4$ (ion-pair energy) and the total free energy of complexation ($\Delta F$), in units of $(N_1+N_2)k_BT$, and are plotted as functions of charge density (ionizability) $f_{m1}$ for (a) lower Coulomb strength $\tilde{\ell}_B=2$ and (b) higher Coulomb strength $\tilde{\ell}_B=2.5$, close to the critical line of $\tilde{\ell}_B^\star$. The parameters are: $N_1=N_2=1000, \delta_1=\delta_2=\delta_{12} \equiv \delta =3.5, w_{ij}=0.0, \tilde{c}_s=0.0$, and $\tilde{\rho}_i=0.0005$. The two $\tilde{\ell}_B$ values are chosen such that the system falls in either side of the crossover Coulomb strength $\delta \widetilde{\ell}_{B}^\star$ for moderate values of $f_{m1}=f_{m2}$. Trends of dominance of enthalpy or entropy of complexation do not change significantly with ionizability.}%
		\label{fig:lbc}%
	\end{figure}

From Fig. \ref{fig:lbc}(a)-(b) we find that with a slight change in $\widetilde{\ell}_{B}$, on either side of the crossover strength $\widetilde{\ell}_{B}^\star$ for a fixed $\delta$, there occurs a role reversal in the drive for complexation, shifting from that of the gain in energy of ion-pairs to the gain in entropy of released counterions.  It has been discussed how this model confirms that both enthalpy and entropy are supportive of complexation at low Coulomb strengths, with enthalpy being the dominant force. At higher Coulomb strengths, enthalpy becomes opposing but entropy continues to be supportive and dominant to drive the process. However, the interplay of enthalpy and entropy is expected to be marginally dependent on the ionizability ($f_{mi}$), unlike the Coulomb strength, as discussed above. To this end, two Coulomb strengths are chosen on either side of $\tilde{\ell}_{B}^\star$ ($\sim 2.25$) at a fixed $\delta=3.5$ (zero salt). At the value of the electrostatic strength lower than the crossover [$\tilde{\ell}_{B}=2.0$, Fig. \ref{fig:lbc}(a)], enthalpy dominates the drive for all ionizabilities, with the dominance being lower for lower $f_{mi}$. At a value of $\tilde{\ell}_{B}$ higher than the crossover [$\tilde{\ell}_{B}=2.5$, Fig. \ref{fig:lbc}(b)], enthalpy is marginally supportive of complexation for lower ionizabilities, and gradually becomes marginally opposing (slight positive change) for higher ionizabilities. The entropy is always favorable to complexation, for all values of $\tilde{\ell}_B$, and for all ionizabilities. Therefore, features of the enthalpy or entropy of complexation (whether it is supportive or opposing) do not change significantly with ionizability, but do with Coulomb strength. The free energy gain increases linearly with ionizability of the polyions for fixed Coulomb strengths.   

\subsubsection{Free energy gain, overlap, and the charge density}

\begin{figure}[!htbp]%
	\centering%
	\includegraphics[height=5.25cm,width=7.00cm]{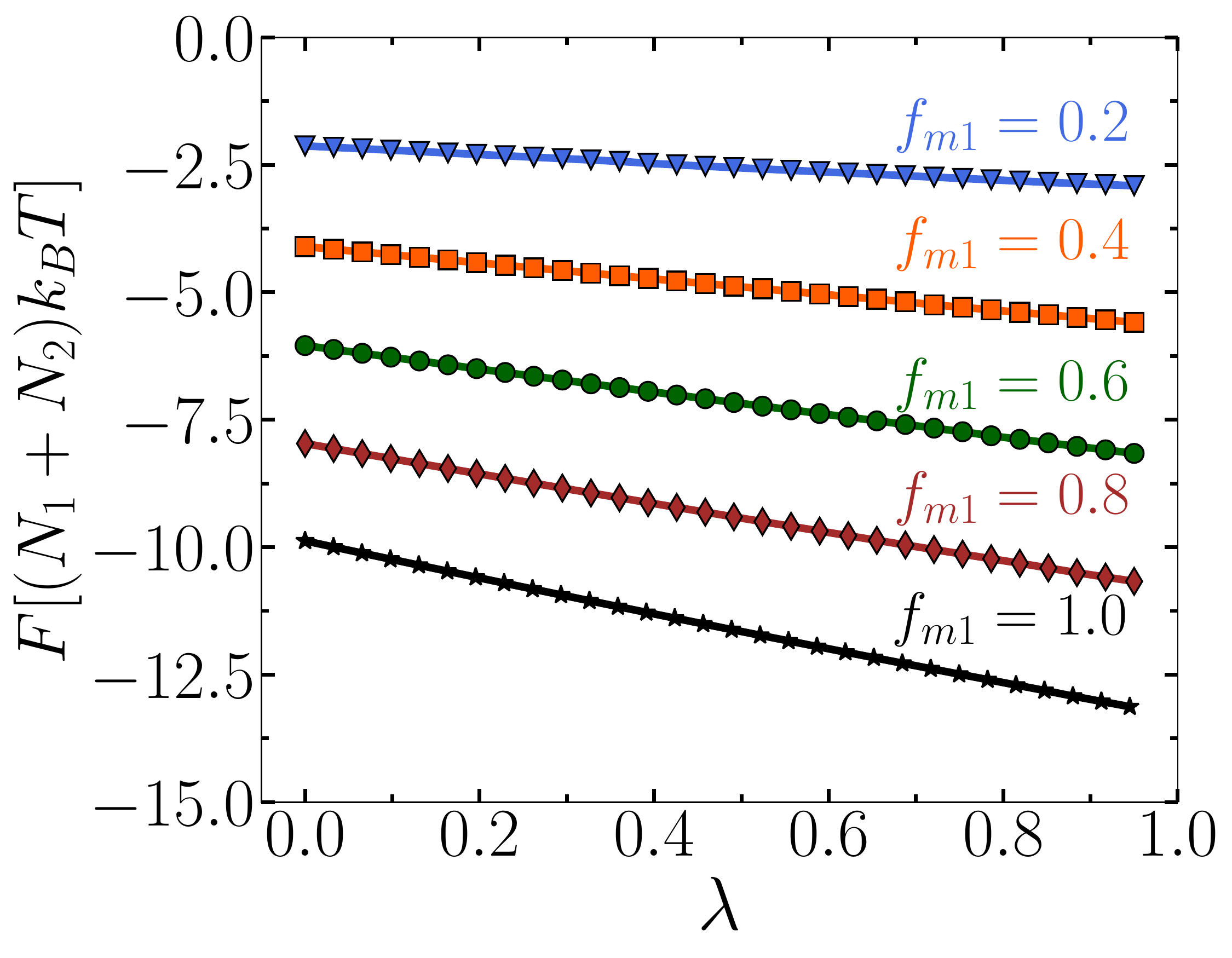}%
	\caption{Free energy versus overlap of complexing polyions: Free energy of the complex, 
	$F$, in units of $(N_1+N_2)k_BT$, is plotted as a function of the overlap of the polyions (length and ionizability of the polyions are taken to be the same), $\lambda=n/N_{c1}=n/N_{c2}$, for different values of ionizability ($f_{m1} \equiv N_{c1}/N_1=f_{m2} \equiv N_{c2}/N_2 =0.2,0.4,0.6,0.8,1.0$). The parameters are: $N_1=N_2=1000, \widetilde{\ell}_{B}=3.0, \delta_1=\delta_2=\delta_{12} \equiv \delta =3.0, w_{ij}=0.0, \tilde{c}_s=0.0$ and $\tilde{\rho}_i=0.0005$.  The curves are linear and decreasing with overlap, showing free energy gain in complexation, which increases with ionizability, for these modest values of Coulomb strength.}%
	\label{fig:freenvsdensity}%
\end{figure}

If the complexation is overall favorable for the pair of two oppositely charged poyions, or, in other words, the free energy of complexation $\Delta F$ is negative (gain), then one may explore the dependency of the free energy on the overlap parameter of the chains. In Fig. \ref{fig:freenvsdensity}, the total free energy per monomer [$F/\{(N_1+N_2) k_B T\}$] is plotted as a function of overlap parameter $\lambda=n/N_{c1}$, the ratio of the number of ion-pairs in the complexed part (at an intermediate degree of overlap) to that in the final complex. For simplicity, both the sizes and ionizabilities of the chains are taken to be equal with $N_1=N_2=1000$ and $f_{m1}=f_{m2}$, leading to $\lambda=n/N_{c1}=n/N_{c2}$. For these intermediate values of the Coulomb strength ($\tilde{\ell}_B=3.0$. $\delta=3.0$, and $\delta \tilde{\ell}_B=9.0$), the free energy is a monotonically and almost perfectly linearly decreasing function for all values of the ionizability. The free energy gain increases with ionizability, which is expected because this is the regime in Coulomb strength for which the enthalpy of complexation is marginally varying, and the entropy gain due to released counterions drives the process\cite{mitra2023} [see Fig. \ref{fig:size}(d)]. As the number of counterions (both initially condensed and finally released) increases with ionizability, so does the free energy gain for complexation. The dependency of $F$ on $\lambda$ as observed, for different values of $f_{mi}$,  is consistent with the results of fully ionized chains described with a slightly different model\cite{mitra2023}, and also with simulations\cite{dzubiella2016,peng2015}. Further, the results are supported by previous simulations and experiments which observe more stable complexes or coacervates for higher charge densities of polyions\cite{juarez2015,kayitmazer2015,huang2019,neitzel2021}, and solid or liquid-like PE complexes based on, respectively, high or low charge densities which dictate the number of released counterions, related entropy gain, and free energy gain or binding strength\cite{chen2021} of the complex.

In the context of the ionizabilities of the polyion chains, two issues related to the free energy gain and associated enthalpic and entropic drives in complexation are worth a discussion, 
First, the dimensionless Bjerrum length $\tilde{\ell}_B=\ell_B/\ell$ (which is the Coulomb strength parameter for fixed 
$\delta$) can be interpreted both in terms of the Bjerrum length $\ell_B$ and the normalizing length
$\ell$, that is chosen as the Kuhn length (length of a monomer for a fully flexible polymer) here. A lower $\tilde{\ell}_B$, one may note that, can occur due to either a lower $\ell_B$ or higher $\ell$, independently. 
We have often interpreted that at low Coulomb strengths (low $\delta \tilde{\ell}_B$'s), or equivalently, at high temperatures or dielctric constants (low $\ell_B$'s), or for bulkier counterions and relatively higher local dielectric constant (low $\delta$'s), the complexation is driven by enthalpy (Fig. \ref{enthalpy-entropy}). Often an alternative explantion is proposed\cite{zhaoyang2006,fu2016,whitmer2018macro}, that low $\tilde{\ell}_B$ occurs due to increasing value of $\ell$, which is the charge separation along the contour of the chain (equivalently, decreasing charge density), keeping $\ell_B$ fixed. Thereafter it is interpreted that, as for lower $\tilde{\ell}_B$'s the complexation drive is enthalpic, it is so for weakly charged systems too. We believe this interpretation may not apply here, because $\ell$ is actually the Kuhn length or the length of a monomer for flexible polymers, whereas the charges can be separated by several, intervening uncharged monomers (several Kuhn lengths) to make the polyion weakly charged (as described by the partially ionizable chains used in this work). Hence, the Coulomb strength will not necessarily be low for weakly charged systems. If only the monomer becomes larger, $\tilde{\ell}_B$ will decrease for the same temperature, but $\delta (=\epsilon \ell/\epsilon_l d$, see discussion after Eq. \ref{F4_papc}) will be higher to keep the effective Coulomb strength $\delta \tilde{\ell}_B$ approximately the same. In such a situation, although the dimensionless results will not change (as $\ell$ is just a scaling factor for length), the interpretation of molarity of the salt, plotted as $\tilde{c}_s=c_s \ell^3$, will. Therefore, as it can be confidently stated that the complexation drive is enthalpic for low Coulomb strengths, the same can not be immediately ascertained for weakly charged (sparsely ionizable) polyions, which can still enjoy high Coulomb strengths that lead to entropy-dominated complexation drives.

Second, there are other situations, especially in simulations\cite{dzubiella2016}, in which Manning condensation is applied to the polyions, leading to fractional charges (a fraction of electronic charge) of monomers. In the case of both monovalent monomers and counterions, Manning argued that the condensation of the counterions occurs if $\ell_{B}\ge \ell$, leading to a condensation threshold of $\widetilde{\ell}_{B}=1$. The infinite dilution assumption validates the consideration of extended rodlike conformation of the PEs. Such mean field assumption of line charge density
in polyions, leading to fractional charges of monomers, ignores the correlations among integer-charged segments, and may lead to erroneous interpretation that the major driving force of complexation in weakly charged polyions is the enthalpy gain. However, as one moves away from infinite dilution and starts exploring more experimentally realistic situations, one must consider the actual distribution of the integer-charged monomer segments, whereas the assumption of a mean-field line-charge density leading to fractional charges for monomers may not be appropriate.

	\section{Conclusions}

We present a general theoretical framework describing the complex formation of two flexible, oppositely charged polyelectrolytes of asymmetric length and partial ionizability (chemical charge density) with neutralizing counterions, by constructing the Edwards Hamiltonian that consists of intra- and inter-chain Coulomb and excluded volume interactions between the segments of the two polymer chains. With the assumption of maximal ion-pair formation in the complexed part of the polyions, the free energy is derived using a variational extremization of the free energy by a simpler trial Hamiltonian, and it explicitly accounts for the screened Coulomb interaction between pairs of distant ions and Coulomb energy of bound ion-pairs. In addition, the free energy comprises entropic contributions from free and condensed counterions, along with the configurational entropy of the polyions. The effective charge (resultant chemical charge after counterion condensation) and size of the complex formed of the polyions are determined by minimizing the free energy that is parametrized by the degree of asymmetry in length and ionizability of the polymers. The thermodynamics of complexation (complexation drive) in the form of free energy gain, and its major enthalpic and entropic components at various electrostatic strengths (or Coulomb strengths, which is a product of Bjerrum length and the dielectric parameter accounting for the local dielectric constants and ion sizes in this model), have been explored in terms of the asymmetries. It is an implicit solvent model that ignores the solvent degrees of freedom, which if considered may significantly change the thermodynamic interpretation at low Coulomb strengths.   	
	
The major features applicable to a fully ionizable PE are observed for the partially ionizable chain as well. Free counterions lead to higher effective charge and size of the chain at moderate, but counterion condensation leads to significant reduction in both at high, Coulomb strengths.  However, as expected, there are decreased electrostatic effects with lower ionizabilities. The theory reproduces known and previously understood trends for complexation of fully ionizable, symmetric, oppositely charged polyions. The free energy drive, as expected, is dominated by enthalpy of bound ion-pairs and entropy of released counterions at low and high Coulomb strengths, respectively.    	
	
For the PE complex made of fully ionizable polyions of asymmetric lengths, only a part of the charge of the larger chain remains compensated; hence the complex remains still expanded. The effective charge and size of the complex decrease with higher symmetry. The free energy gain per monomer increases monotonically with increasing degree of symmetry, reaching the maximum and corresponding to the most stable complex at equal lengths. An asymmetric, soluble PE complex with net charge is expected to repel similar complexes in solution to hinder complex coacervation, the effect decreasing with salt screening, observed in experiments.

For the pair of partially ionizable polyions, we explored a combination of chain sizes and ionizabilities keeping fixed the number of ionizable monomers (the stoichiometric charge content) of the individual chains. The asymmetry in ionizabilities between two complexing polyions with symmetric lengths produces results very similar to the pair with symmetric charge densities but asymmetric lengths. Such agreement for both the effective charge and size of the complex implies the significance of the amount of stoichiometric charge of the chains, less of their size and ionizability, in PE complexes.  In both cases, an increasing degree of asymmetry increases the effective charge and size of the complex, which decrease in the presence of salt that screens almost all electrostatic effects. The trends predicting the correct premises for coacervation match with simulations and experiments.   

For complexing polyions of the same length and ionizability, the latter varying, similar trends of dominance in the drive, changing from being enthalpic to entropic demarcating the electrostatic phase space into two regimes, is observed. The magnitudes of the drive, which are of electrostatic origin, decrease with lower charge densities, but, surprisingly, the crossover Coulomb strength separating the regimes is marginally sensitive to it, because the degree of condensation counterions (considering only ionizable monomers), and the consequent release of them which dictates the crossover from enthalpic to entropic regime, is found to be insensitive to the polyion charge density.  The crossover Bjerrum length decreases with the dielectric mismatch parameter and salt, due to increased counterion condensation. However, the product of the Bjerrum length with either of the two (keeping the other fixed) is almost constant, because the fraction of condensed counterions is almost uniquely determined by the product.
 
The enthalpy of complexation changes from being favorable to opposing, for a slight change on either side of the crossover Coulomb strength for a fixed salt (here, zero salt). However, for the entire range of ionizability of the polyions, the enthalpy change, whether positive or negative, remains marginal at this crossover strength. The entropy and free energy gain, however, are seen to increase monotonically with ionizability for all Coulomb strengths. This shows why the interplay of enthalpy and entropy, and the trends of their relative dominance, do not change sensitively with ionizability. 

The total free energy as a function of overlap of the complexing polyions, for moderate choices of Coulomb strengths, is seen to decrease monotonically and linearly, rendering the process favorable for complexation. This trend matches with simulations. Higher ionizability results in a larger number of released counterions, which increases the entropy gain, leading to a more favored complex formation. Since modest Coulomb strengths (that is the mid-range values in our work) witness a marginal enthalpy loss, the total free energy is handsomely driven by the counterion entropy exclusively, thus providing a larger free energy gain of complexation for higher ionizability. 

\begin{section}{acknowledgement}

The authors acknowledge financial support from IISER Kolkata, Ministry of Education, Government of India. They also thank Aritra Chowdhury and Ben Schuler for discussions that led to a better understanding of the molecular interactions in the system.

\end{section}

	\bibliography{partialCOMP}
	
\end{document}